\documentclass{optica-article}

\journal{opticajournal} 

\articletype{Research Article}

\usepackage{lineno}

\begin{document}

\title{Astrophotonics - current capabilities and the road ahead}

\author{Barnaby Norris,\authormark{1,2,3}, Simon Gross,\authormark{4} Sergio G. Leon-Saval\authormark{1,2,3}, Christopher H. Betters\authormark{1,2,3}, Julia Bryant\authormark{1,2,3}, Qingshan Yu\authormark{1,2,3}, Adeline Haobing Wang\authormark{1,2,3}, Glen Douglass\authormark{4}, Elizabeth Arcadi\authormark{4}, Ahmed Sanny\authormark{4}, Michael Withford\authormark{4}, Peter Tuthill\authormark{1,2,3}, Joss Bland-Hawthorn\authormark{1,2,3}
}

\address{\authormark{1}Sydney Institute for Astronomy, School of Physics, University of Sydney, Australia\\
\authormark{2}Sydney Astrophotonic Instrumentation Laboratories, School of Physics, University of Sydney, Australia\\
\authormark{3}Astralis, Astronomical Instrumentation Consortium, Australia\\
\authormark{4}MQ Photonics Research Centre, Macquarie University, Sydney, Australia}

\email{\authormark{*}barnaby.norris@sydney.edu.au} 


\begin{abstract*} 
Astrophotonics represents a cutting-edge approach in observational astronomy. This paper explores the significant advancements and potential applications of astrophotonics, highlighting how photonic technologies stand to revolutionise astronomical instrumentation. Key areas of focus include photonic wavefront sensing and imaging, photonic interferometry and nulling, advanced chip fabrication methods, and the integration of spectroscopy and sensing onto photonic chips. The role of single-mode fibres in reducing modal noise, and the development of photonic integral field units (IFUs) and arrayed waveguide gratings (AWGs) for high-resolution, spatially resolved spectroscopy will be examined. As part of the Sydney regional-focus issue, this review aims to detail some of the current technological achievements in this field as well as to discuss the future trajectory of astrophotonics, underscoring its potential to unlock important new astronomical discoveries.

\end{abstract*}

\section{Introduction -- moulding the flow of starlight}
In its constant quest to push back the limits of the observable universe, astronomy has long sat at the bleeding edge of technological innovation. In some cases technical developments conceived to solve the near-impossible tasks demanded by astronomical science have made their way into mainstream applications, while other times the astronomical community has excitedly adopted developments from industry as a means to tackle the latest astrophysical questions. Either way, major goals such as making images of Earth-like planets around distant stars, or glimpsing the primordial light of our universe emitted fractions of a second after its creation, hinge on rapid adoption of innovation in astronomical instrumentation.

In terms of optical (as opposed to radio) astronomy, nowhere has this been more pronounced than the impact of photonic technologies. Catalysed by the initially communications-driven revolution in photonic technologies, astrophotonics is already changing the landscape of astronomical instrumentation. Photonics has the potential to revolutionise the achievable performance and fundamental types of observations possible in optical astronomy. But while this has already made important inroads in some areas, with high profile results (such as the observation of the galactic centre with the GRAVITY instrument \cite{GRAVITY2018}), the adaptation of photonic devices to the unforgiving domain of astronomy is extremely challenging, as discussed in Section \ref{sec_challenges}. 

Fortunately, the initial successes of astrophotonics has generated considerable, rapidly increasing momentum in the astronomical community, with many of the world's leading astro-instrumentation researchers working on tackling these challenges to ensure a successful realisation of photonic technologies in current and future instruments\cite{Jovanovic2023}. 
The state of the art of astrophotonics encompasses on-sky instruments ranging from technology demonstrators to science powerhouses in addition to ongoing laboratory, fabrication and design work. Together, these aim to conceive the next generation of integrated astronomical instruments, unlocking previously impossible observational domains and, in particular, being crucial to realising the performance of the coming generation of Extremely Large Telescopes (ELTs). 

In this review paper, published as part of the Sydney regional focus issue of Applied Optics, we present an overview of the current state of a range of innovative astrophotonic technologies, along with challenges and potential next steps needed to realise the widespread use of these methods. Astrophotonic devices for spectroscopy, interferometry, wavefront sensing and imaging are discussed, with a particular focus on those technologies being pursued most actively by Sydney researchers.

\section{Astronomy's unique challenges for photonics}
\label{sec_challenges}

While exploiting photonic technologies enables an impressive range of observational possibilities, there are considerable challenges faced in astronomy which are not necessarily encountered in other photonic applications. The two largest ones are the extremely low light levels involved, and the distortion caused as starlight passes through the Earth's turbulent atmosphere (\emph{seeing}).

The Earth's atmosphere includes multiple layers of turbulent air, with air of differing temperature (and refractive indices) mixing chaotically and driven laterally by the wind, evolving on millisecond timescales. Seeing -- the effect of this on starlight as it transits the Earth's atmosphere on the way to a terrestrial telescope -- is one of the main limiting challenges in optical astronomy today whenever high spatial resolution, or efficient injection into a photonic device, is required. To put this in perspective, the diffraction-limited angular resolution of a current generation 8~m diameter telescope at visible wavelengths is approximately 15 milli-arcseconds (mas), and the 39~m diameter of the coming Extremely Large Telescope has a diffraction-limited resolution of approximately 3~mas. This puts the direct imaging of nearby habitable-zone planets, which have star-planet separations of order 10~mas, within reach. However, the blurring effects of atmospheric seeing at a typical observatory limits the achievable resolution to approximately 1000~mas!
As detailed in Section \ref{sec_pl}, this seeing-induced limit on spatial resolution equivalently restricts the ability to inject starlight into single-mode or few-mode fibres and photonics. Various approaches are being used, and continually being developed, to tackle this problem.

Another key challenge is that, unlike in photonic domains using a laser or other laboratory light source, astronomy is almost always operating in the photon-starved regime. By the very nature of the quest to reach more distant and more compact targets, the limiting factor in observations is quite often the total amount of light that can be collected during an observing period (underscoring the importance of high injection efficiencies as discussed above). This has driven the construction of larger and larger telescopes to increase the collecting area, including the coming generation of Extremely Large Telescopes (ELTs). These include the Giant Magellan Telescope (GMT)\cite{Johns2012} with a 25~m diameter primary mirror, the Thirty Metre Telescope (TMT)\cite{Sanders2013} with a 30~m mirror and the ESO Extremely Large Telescope (ELT)\cite{Gilmozzi2007, Tamai2022} with a massive 39~m diameter mirror. The cost of building such massive, steerable mirrors with nanometre-level surface accuracy is extreme, and the larger apertures only increases the challenge of recovering their diffraction-limited performance and thus efficiently injecting light into single-mode waveguides \cite{Guyon2005}. 

This places extra requirements on photonic devices. Device throughput is paramount, with insertion losses of even just 1~dB or so being intolerable, restricting device design and complexity, and limiting access to some photonic platforms. For example a prototype device for a future upgrade to the FIRST interferometer\cite{Huby2012} -- which performs complicated phase modulation and beam combination -- had very low throughput (1-2\% at 750~nm)\cite{Martin2022}, partly due to the mode-field mismatch between the electro-optic phase modulation chip (using LiNbO$_3$ chip) and the beam recombination chip (silica). Excellent progress is being made in improving this throughput\cite{Lallement2023} through new designs and processes. In general, throughput losses encountered when high index-contrast photonic platforms (such as LiNbO$_3$) are coupled with the standard single-mode fibres used at observatories is an important problem, as is the different geometric mode-field shape between SMF and rectangular waveguides in planar photonic devices. For devices made using Ultrafast Laser Inscription (ULI) (see Section \ref{sec_uli})
the low index contrast strongly limits bend radii (and increases device length), though techniques such as annealing can help maximise throughput\cite{Norris2014}.

If a photonic device which operates only in one polarisation is used, removing one polarisation by passing it through a linear polariser (and thus losing 50\% of the starlight) is unacceptable, and so either polarisation-insensitive devices must be used or the entire photonic system duplicated, one for each linear polarisation, and starlight distributed between them via a polarised beamsplitter along with complex polarisation control components, such as in the GRAVITY instrument\cite{GRAVITY2017}. And while the polarisation dependence of photonics has often been treated as a nuisance, astronomers are now looking to perform precise polarimetric imaging using photonic devices\cite{GRAVITY2023}.
The low flux levels also preclude access to many of the non-linear optical techniques enjoyed in laboratory settings. A related challenge is the requirement for very high contrast measurements in astronomy (for areas such as exoplanet imaging), where cross-talk of just 1 part in $10^6$ or even $10^9$ can be ruinous due to stellar photon noise contamination. 

Compared to what is usually used for other photonic applications, large bandwidths -- motivated largely by the requirement to utilise all available starlight -- are required for astronomical purposes. For example a basic requirement for a photonic instrument operating in the standard astronomical H band (often the first choice for astrophotonic devices since it overlaps with the telecom C-band, leveraging the existing technology development there) would be to function over the bandpass from 1.5$\mu$m to 1.8$\mu$m. Compare this to the telecom C-band width of 1.53$\mu$m to 1.565$\mu$m, and it is clear development beyond existing communications-focused devices is required, and is the focus of much current research. 

In the subsequent Sections, the ways in which these technologies are being adapted to suit astronomical demands will be outlined, and exciting applications and results achieved thus far examined. Challenges and possible next steps will also be explored, along with a discussion of the potential scientific boons that may be realised. 

\section{Photonic lanterns for wavefront control, optimal injection and imaging}
\label{sec_pl}
A rapidly growing astrophotonic technology in recent years is the photonic lantern, and its use for wavefront sensing, imaging and highly efficient single-mode fibre injection. A photonic lantern (PL)\cite{Leon-Saval2005, Leon-Saval2013, Birks2015} is a device that efficiently transforms light from a multimode fibre into multiple single-mode (SM) fibres. This requires the number of output SM fibres to at least match the number of modes in the multimode input to ensure high efficiency conversion. 

This conversion is enabled by a specially engineered taper transition. In this transition, the cores of the SM fibres gradually vanish, and their cladding transforms into the core of the newly formed multimode fibre. The cladding of this new fibre is composed of a low refractive index capillary. These devices were initially manufactured by placing individual SM fibres into a capillary and tapering them. More recently, the use of multi-core fibres (MCF) has become common\cite{Birks2012}. MCFs, which contain multiple separate single-mode cores, replace the need for multiple bulk fibres, simplifying the manufacturing process and enabling the production of devices with a greater number of modes/cores. Alternatively, integrated photonic lanterns can be produced using the laser direct write process\cite{Spaleniak2013}, as described in Section \ref{sec_uli}. 
The loss from the multi-mode to single-mode transition of a standard tapered PL has been measured to be $\sim$0.3~dB, with a similar value ($\lessapprox$0.5~dB) for multicore-fibre based devices \cite{Birks2015}. Laser direct-write based devices have demonstrated losses of $\sim$1~dB\cite{Spaleniak2013}, though these are at a less mature stage of  development than fibre-based devices. 

Initially developed to operate around the astronomical H band\cite{Leon-Saval2005}, in principle a photonic lantern can be developed for any given wavelength, given suitable materials and fabrication techniques. The silica fibre based techniques initially used have been successfully used to produce PLs operating at visible wavelengths \cite{Norris2020}. But building efficient devices for wavelengths longer than $\sim$2~$\mu$m requires new materials which are transmissive at these wavelengths, such as fluoride optical fibers. ULI-based PLs may be produced using the demonstrated ability of this process to produce mid-IR capable devices, operating at wavelengths as long as 4~$\mu$m, using chalcogenide glasses \cite{Gretzinger2019}.

\begin{figure}[h]
\centering\includegraphics[width=0.8\textwidth]{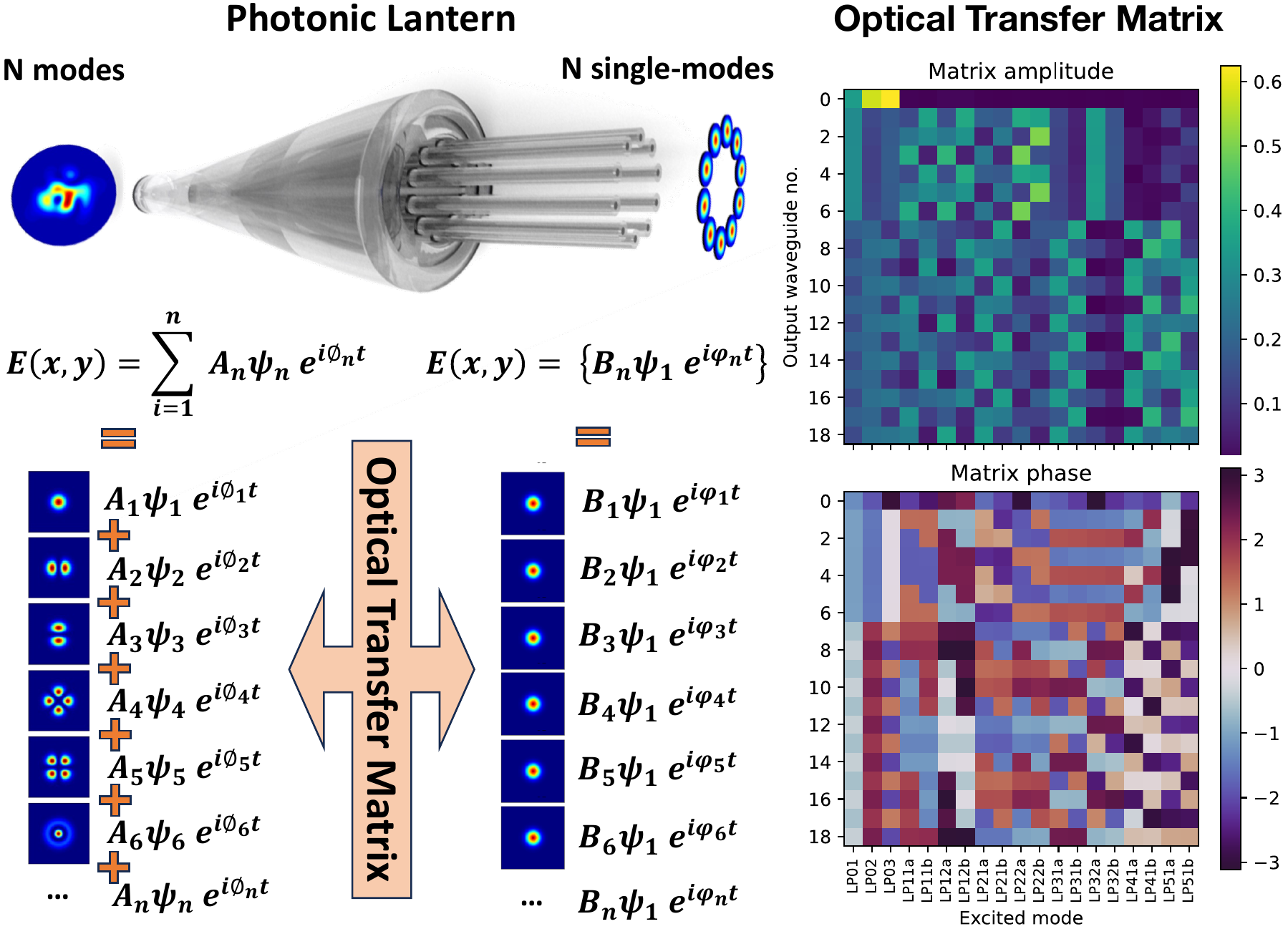}
\caption{Conceptual diagram of the photonic lantern. When used as a wavefront sensor, the aberrated PSF is injected into the multimode end and excites the N modes supported by the photonic lantern. This is mapped to the amplitudes of the N output single-mode fibres according to the complex transfer matrix of the device. Given this matrix the incident PSF amplitude and/or phase can be reconstructed from the measured single-mode fibre fluxes. In practice this transfer matrix is very stable but must be measured after the device is fabricated.}
\label{fig_plmatrix}
\end{figure}

In a typical PL, the output from each fibre is not a direct mapping of each input mode. Instead, the complex output of each fibre is a linear combination of the complex amplitudes of all input modes. An example of this relationship is shown in Figure \ref{fig_plmatrix}, which includes the complex transfer matrix for a 19 core, 19 mode device (built from a MCF  with hexagonally-arranged cores) operating at 1550 nm, modelled using RSoft. Although such matrices can be modelled, acquiring them for a physical device is generally not feasible. In practical applications, the intensity of the output fibres, which is the square of the complex amplitude, is measured. This introduces a \emph{nonlinear} relationship between the measured SMF intensities and the wavefront phase and/or amplitude\cite{Norris2020, Wong2022}. Another type of PL -- the mode-selective PL\cite{Leon-Saval2014} -- operates differently. In this case, a given output fibre corresponds directly to a given input mode, rather than being mixed via a transfer matrix. These utilise physical modifications to the geometry of the PL, though are limited in the maximum number of modes they can operate over\cite{Velazquez-Benitez2018}. 

The photonic lantern was originally developed for fibre-fed spectroscopy to split a multimode fibre into a set of single-mode fibres, to mitigate modal noise and to allow its use with arrays of single-mode photonic devices. But in those applications the actual relationship between the spatial distribution of light at the multimode input and the distribution of flux in the single-mode outputs was ignored. The more recent excitement around PLs arises from methods of \emph{utilising} this encoded spatial information, as this encoded information contains crucial types of signal that can not be detected by a standard imaging detector.

Specifically, the excitation of the different modes in the multimode fibre
is a direct function of both the \emph{amplitude} and the \emph{phase} of the injected image (its complex electric field). In contrast, a conventional camera can only measure the light's intensity (the square of the amplitude), not the phase. A photonic lantern is able to convert the complex values of each of these input modes into the distribution of fluxes in a set of single-mode fibres. Measuring the output flux of a set of single-mode fibres is simple, and measuring them as a function of wavelength is also straightforward (as is done in any fibre spectrograph).

Thus, as long as the photonic lantern's transfer function  -- mapping the complex electric field of the input image to the flux values of the output fibres -- is known, then one can work backwards and reconstruct the complex input electric field from the measured fibre fluxes. While it is difficult to precisely prescribe this exact mapping before manufacture, once the PL is made this transfer function is extremely stable, as the PL is a single solid monolithic glass device. The PL's transfer function can then be measured, and then that one-off calibration stored and reused indefinitely for wavefront or image reconstruction. Since this transfer function is non-linear (because the measured intensity is the square of the complex amplitude), a non-linear algorithm such as a neural network is used to reconstruct the input field. Characterising the long term stability of the PL's transfer function is a key focus of current research. In experiments thus far, stability measurements have instead been dominated by drifts in the bulk optical and mechanical mounting components of the experimental setup, and so further work -- either in precisely environmentally controlled laboratories or via some type of differential measurement -- are required. This rapidly developing technology is now beginning to enable revolutionary new applications, as discussed in the following subsections.

\subsection{Focal plane wavefront sensing}
High resolution and high contrast imaging in astronomy generally relies on adaptive optics\cite{Davies2012, Guyon2018} (AO), where a wavefront sensor (WFS) is used to measure the instantaneous wavefront error induced in the incident light and then rapidly correcting it by applying the inverse of the wavefront error to a deformable mirror (DM). This must happen at kilohertz rates as the atmospheric phase structure changes on millisecond timescales. Usually a wavefront sensor placed at the pupil plane, such as a Shack-Hartmann \cite{Platt2001} or pyramid wavefront sensor \cite{Ragazzoni1996} is used. However these have some significant limitations. One is that they suffer from non-common-path aberrations (NCPA) -- the wavefront seen by the WFS is different to that which creates the science image due to the non-common optical components traversed by these two beams\cite{Sauvage2007}. Another is that they are not able to sense certain `blind' modes -- extremely damaging aberrations involving phase shears which arise from the telescope structure, as described below.

Thus there is strong demand for a \emph{focal plane} wavefront sensor (FP-WFS), driven by an ability to detect blind modes and being located at the same optical plane and wavelength as the science image, avoiding non-common-path aberrations. Blind modes - in particular low wind effect (LWE) and petalling modes\cite{Sauvage2016, Milli2018, NDiaye2018, Vievard2019} are currently a major problem in AO. LWE arises primarily from phase discontinuities created across the secondary-mirror supports of the telescope, known as spiders. These discontinuities are often intensified by thermal effects in low wind conditions, leading to a significantly disrupted point spread function (PSF). A similar effect is anticipated to occur from imperfect phasing between the mirror segments of the future Extremely Large Telescope (ELT) generation of telescopes, and must also be sensed and corrected.
Traditional pupil-plane wavefront sensors, such as pyramid and Shack-Hartmann sensors, fail to detect this effect. However, it is readily observable in the focal plane.

One of the complexities of using the focal plane image directly as a WFS is that intensity information alone cannot resolve the sign of an aberration, necessitating a method to measure both the phase and amplitude of the electric field at the focal plane. Various approaches to address this challenge have been proposed\cite{Mocoeur2009, Vievard2018, Korkiakoski2014, Martinache2013, Martinache2014}, each with its limitations. These methods often rely on a linear approximation, assuming a wavefront error (WFE) less than 1 radian, which may not hold true for LWE. Additionally, they may require iterative processing, which is not viable for real-time applications, active modulation of a deformable mirror (DM), or alterations to the visible pupil.

\begin{figure}[h]
\centering\includegraphics[width=1\textwidth]{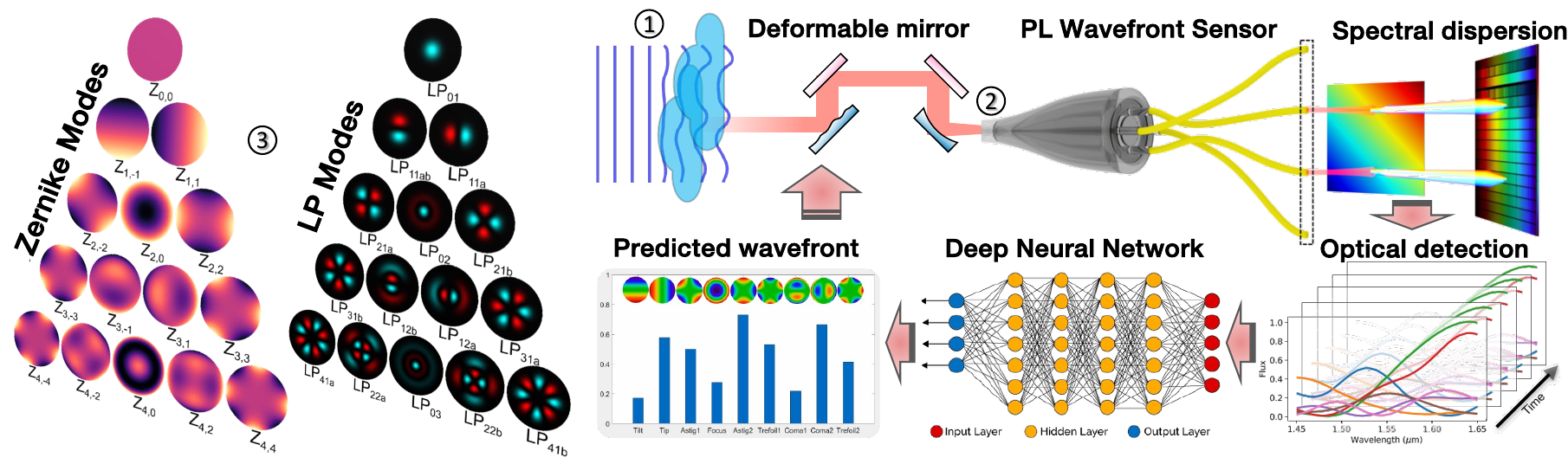}
\caption{An overview of the photonic lantern wavefront sensor. Starlight, after suffering phase aberrations from the atmosphere, is injected into the multimode end of the photonic lantern. Here the complex electric field of the PSF is mapped onto the modal basis (shown here as ideal LP modes) of the multimode region. The photonic lantern then transitions this to a set of single-mode fibre outputs, the amplitudes of which encode the complex spatial information of the original PSF. After output fluxes are measured (with spectral dispersion if desired) a neural network (or other algorithm) is used to reconstruct the wavefront error that was induced by the atmosphere, in any desired modal basis (Zernike modes shown here). The inverse of this (weighted by some gain factor <1) is then applied to the deformable mirror to correct the wavefront.}
\label{fig_plwfs}
\end{figure}

This motivates the development of a photonic lantern wavefront sensor (PL-WFS)\cite{Norris2020}, as outlined in Figure~\ref{fig_plwfs}. When placed at the focal plane a photonic lantern is ideal at sensing both the phase and amplitude of the telescope PSF, since (as described above) the incident electric field can be reconstructed from its easily measurable single-mode outputs. It also allows optimal use of detector pixels (one pixel per spatial mode per wavelength channel) to minimise read noise, it is trivial to spectrally disperse and perform multi-wavelength wavefront sensing to unwrap phase and sense scintillation, the entire device fits inside a regular fibre connector (such as an FC/PC or SMA connector) simplifying deployment and use in multi-object AO systems, and it is ideal for truly zero-non-common-path correction for single-mode fibre injection).

A key advantage of this approach is it mitigates issues arising from any time-varying mode scrambling (e.g. due to strain, temperature) that would occur as multimode light is propagated via multimode fibres. Instead, since the PL is positioned directly at the telescope focal plane, the only multi-mode region is the beginning of the photonic lantern taper itself, which is a few-millimetre-long portion of rigid glass encased in the metal fibre connector. All coherent, phase-dependent components of the signal are encoded here, and subsequent transmission is via sets of independent single mode fibres, of which just the intensity is measured. 

The photonic lantern wavefront sensor (PL-WFS) and wavefront reconstruction via a neural network has now been demonstrated in laboratory settings with excellent low-order, wavefront reconstruction results including for low-wind effect modes\cite{Norris2020,Norris2022b}. Detailed theoretical frameworks, models and design constraints have been developed\cite{Lin2021,Sweeney2021,Lin2022}, including demonstration of the ability to use linear or quadratic reconstruction algorithms for lower wavefront error regimes\cite{Lin2022c}. Laboratory simulations have explored coupling efficiency and identified next steps for improvement\cite{Lin2022b} and even designs which optimise coupling for an off-axis science target and minimise that of a central star have been explored\cite{Xin2022}.  

Programs to implement and evaluate the use of PL-WFS on various major observatories are now underway. A key future development will be the addition of more modes and output SM cores. Current devices have 10s of modes and so are useful as low-order wavefront sensors (which is ideal, as this is where the vast majority of wavefront error occurs), but for high contrast images 1000s of modes, comparable to that of a modern pyramid or Shack-Hartmann WFS, would be required.

\subsection{Optimal SM injection} 
\label{sec_sminj}
Central to the use of any photonic integrated circuit, fibre beam combiner, photonic spectrograph, fibre Bragg grating, etc. is the ability to efficiently inject starlight into one or more single-mode fibres. But due to the effects of the Earth's turbulent atmosphere (\emph{seeing}) this is extremely difficult to accomplish \cite{Jovanovic2016}. The phase error induced by the atmosphere turns the telescope's diffraction-limited PSF into a large cloud of speckles with complex high-order spatial structure; a very poor overlap with the mode of a single-mode fibre. Spatial structure beyond the roughly Gaussian mode field of the SMF is not injected, nor is the corresponding power, and the injection of uncorrected starlight into a SMF is extraordinarily inefficient. Seeing of just 0.3" on an 8~m telescope would reduce injection efficiency to $\sim$3\%, and 0.6" seeing reduces it to essentially zero\cite{Ellis2021}. One way to counter this is using a standard adaptive optics system to create an as close to diffraction-limited PSF as possible\cite{Jovanovic2016}, which has demonstrated injection efficiencies of $\sim$50\% using the SCExAO extreme-AO system at the Subaru Telescope\cite{Jovanovic2016} following optimal fibre positioning. But this faces the same limitations as any pupil-plane WFS based AO system -- especially susceptibility to blind modes and non-common path aberrations. The latter is especially problematic in this case as the single-mode fibre and PSF must be aligned to within a few microns, and it is easy for the relative tip/tilt and low order aberrations between wavefront sensor and fibre to drift by this much over a short period of time, especially given the hostile thermal and mechanical environment of a telescope. 

A focal plane wavefront sensor, such as the PL-WFS described above, offers a clear advantage. But better still, if the injected starlight can be directed entirely into one of the photonic lantern's SMF outputs then this could be sent to the single-mode science instrument, such that the wavefront sensor and science fibre are all the same device, completely eliminating non-common-path aberration. One option would be to determine the wavefront that would maximally excite one output fibre, then apply this via the system's DM by setting it as the AO system's target wavefront, which has been demonstrated in simulations\cite{Sweeney2021}. But the transfer function of the photonic lantern is highly dependent on wavelength, since it results from the interference of all input modes, and so this method would only work in the monochromatic case.

A promising solution is the hybrid mode-selective photonic lantern\cite{Norris2022c}. Unlike a standard photonic lantern, a specific input mode (such as the LP$_{01}$ mode) is mapped over a broad bandpass to one specific output fibre of the PL. This is accomplished by introducing carefully designed structural perturbations to the PL; in the case of a 19 core LP$_{01}$ mode-selected lantern this can be accomplished by tuning the size of the central core and including spatial filtering. Simulations show this can result in only the LP$_{01}$ mode being coupled to the science output fibres with all other modes suppressed by better than 20~dB across the entire astronomical H band (1.5$\mu$m to 1.8$\mu$m)\cite{Norris2022c}. But unlike a fully mode-selective lantern, the other modes are mixed via a transfer matrix. This is optimum for wavefront sensing, as a fully mode-selective PL is not optimal for this application\cite{Lin2022}, as well as being unable to support a large number of modes.

In operation, all other fibres would be used for wavefront sensing as in the usual PL-WFS, controlling the AO system to maximally excite the LP$_{01}$ mode of the PL. 
One unique control-algorithm challenge is this means that when wavefront correction is optimum the signal/noise ratio of the wavefront sensing measurements is at a minimum (since most light is directed into the science fibre). When the correction solution strays, light begins to appear in the other outputs and the AO system then attempts to eliminate it. In any case, injection into a PL would be far more efficient than a standard single mode fibre -- the multimode input of a PL is the same as a conventional multi-mode fibre, and in the AO-assisted case coupling efficiency can simply be optimised by increasing the number of modes in the MMF and PL. But even in the seeing-limited case the injection efficiency is improved, with a 60~$\mu$m core diameter ($\sim$55~mode) PL having an injection efficiency of $\sim$40\% in 0.5" seeing, up from essentially zero for a SMF\cite{Ellis2021}.

\subsection{Future innovation: ultra high-resolution imaging spectroscopy with photonic lanterns} 
To perform astronomical imaging using the photonic lantern wavefront sensing concepts presented thus far, the PL-WFS would be used as a separate device from the science imaging camera, with light possibly split by a beamsplitter or dichroic mirror as in standard AO systems (but immediately before the focal plane). But the ultimate goal is to perform wavefront sensing and imaging entirely within the photonic lantern itself. If the entire complex electric field and spatial coherence properties of the incident starlight were measured by the photonic lantern, then not only could AO correction be performed but potentially the science image could be entirely disambiguated from the effects of seeing during data analysis, producing a diffraction-limited image even in the face of high wavefront error. Moreover, since the outputs of the PL can be easily spectrally dispersed, it would be possible to reconstruct the science image as a function of wavelength, thus producing spectrally resolved imaging. This would be key to characterising exoplanet atmospheres, detecting accreting planets via emission lines, and other cases.

However, the conventional photonic lantern discussed thus far does not have the measurement degrees of freedom to accomplish this. With $n$ input modes and $n$ SM outputs, and with only the intensities of those outputs being measured, there is not enough information to completely constrain the phase, amplitude, and coherence of the input PSF. This is not a problem in other applications such as wavefront sensing, where the implicit assumption is made that atmospheric seeing modulates the phase but not the amplitude of the wavefront. One possibility is to take the SM outputs of the PL and interferometrically recombine them in a separate interferometry photonic device\cite{Kim2024}, as is done in long baseline interferometry (see Section \ref{sec_interf}). A possible challenge with this method is it requires that the optical path difference between the various single-mode fibres connecting the PL to the beam combiner chip remains completely stable, otherwise the transfer function will change.

Another approach would be to use a so-called oversampled photonic lantern. This would have a greater number of output SM fibres than input modes, allowing measurement degeneracy to be broken. This requires that all output fibres independently sample the interference taking place within the PL taper, which simulations show is possible as long as one is careful to avoid geometric symmetries in construction which can lead to redundancy in the outputs. In theory, to completely constrain the amplitude, phase and spatial coherence (equivalent to fringe visibility in an interferometry context) of the input signal, three outputs would be required for each input mode, though given the advances made in interferometric image reconstruction algorithms it is possible that in reality fewer would be required. It is also necessary to ensure sufficient sampling (and constrain the transfer function model accordingly) to resolve any degeneracy between spectral features of the source object and wavefront-induced chromatic modulation of the PL outputs. This work is in its early stages, with laboratory results from prototype devices now having demonstrated the independent reconstruction of phase and amplitude using such a device\cite{Norris2024ip}.

\section{Photonic interferometry and nulling}
\label{sec_interf}
Arguably the biggest impact of astrophotonics has been in the domain of high spatial resolution and high contrast imaging via interferometry. In astronomical interferometry\cite{Shao1992, Monnier2003, Labeyrie2006} light from multiple telescopes, or from multiple sub-pupils from a single telescope, is coherently combined and the resulting fringe pattern analysed. When combining light from multiple telescopes (long baseline interferometry), such as is done at the VLTI\cite{Haguenauer2012}, CHARA\cite{tenBrummelaar2005}, and other such facilities, the maximum achievable spatial resolution is set not by the diffraction limit of a single telescope (i.e. angular resolution limit $\theta$ being $\sim \lambda/D$, where $D$ is the diameter of the telescope) but by the longest separation between any pair of telescopes (baseline length, $B$), with the angular resolution limit $\theta$ conventionally taken to be $\sim\lambda/(2B)$. 

Beyond the raw increase in diffraction-limited performance, interferometry has another key advantage -- the produced observable quantities (such as fringe visibilities and closure phases) are highly robust against seeing, motivating the use of this technique even on a single telescope using techniques such as aperture masking\cite{Tuthill2000} which divide the telescope's pupil into a set of subapertures. Fringe visibilities are defined as the contrast of a fringe pattern for a given baseline (corresponding to a specific measured spatial frequency), as per 
\begin{equation}
V = \frac{I_{max}-I_{min}}{I_{max}+I_{min}}
\end{equation}
where $I_{max}$ and $I_{min}$ are the maximum and minimum fringe intensities. In practice, $V^2$, derived from the power spectrum (the squared modulus of the Fourier transform of the fringe pattern). Crucially, given the assumption that seeing affects only the phase (not amplitude) of the wavefront, seeing may modulate the location of interference fringes but as long as the detector integration time is less than the atmospheric coherence time $\tau_0$ ($\sim 1$~ms) their \emph{visibility} would not be affected. The goal here is to measure the spatial coherence function $\gamma$ of the observed source, since (as per the Van Cittert-Zernike theorem\cite{Labeyrie2006}) for a distant astronomical object the Fourier transform of its spatial coherence gives its angular intensity distribution, i.e. image. This is directly related to the visibilities, as
\begin{equation}
\label{eqn_vis2coh}
V = \frac{2 A_1 A_2}{A_1^2 + A_2^2} \cdot | \gamma(\mathbf{r}_1, \mathbf{r}_2) |
\end{equation}
where $A_1$ and $A_2$ are the amplitudes of the measured waves. Phase information can also be recovered by the use of closure phase\cite{Baldwin1986}, calculated taking the vector sum of baseline phases around a closed triangle, which ideally will cancel out atmospheric phase errors.

\subsection{Spatial filtering}
In reality things are not so simple, and conventional bulk-optics implementations of these techniques face a number of limitations, but which astrophotonics can resolve. The first major advantage is spatial filtering. The idealised ability of visibilities and closure phase to be unaffected by phase error from seeing as described above assumes each measurement is a single scalar value of the electric field at a given point.
In reality the telescope apertures that are used to form the baselines have pupils of finite size, which each contain complicated phase structure induced by the atmosphere. But by injecting the light from each telescope of a long baseline interferometer into a single-mode fibre or waveguide, all high order spatial structure is removed (at the expense of injection efficiency) and much more accurate interferometric observables can be produced. This method has now become commonplace, from the pioneering FLUOR instrument\cite{duForesto1998} to current leading instruments such as MIRC-X\cite{Anugu2020} and GRAVITY\cite{GRAVITY2017}. Here,  the light from each telescope is injected into a separate single-mode fibre that can then be used to form fringes on a detector in the conventional way, or instead be measured using a photonic chip as described below. 

The same principle is also used in pupil remapping interferometry, and alternative to aperture masking interferometry for a single telescope. In aperture masking interferometry, in order to obtain a unique measurement of each spatial frequency the pattern of sub-apertures must be non-redundant, that is to say the vector separation of each and every pair of holes must be unique. This constraint massively limits the throughput, with greater than 90\% of the telescope pupil often blocked out. Pupil-remapping involves using single-mode fibres or waveguides to rearrange a pattern of sub-apertures into a suitable configuration for recombination, such as a linear non-redundant array, while maintaining temporal coherence. This reconfiguration facilitates the formation of fringes on a detector or the injection into a photonic beam combiner device. Notably, the non-redundancy requirement for the input pattern is eliminated, allowing almost the entire pupil to be utilised with sufficient waveguides, significantly enhancing throughput. 

A vital aspect of this technique is maintaining optical path length coherence, despite the potentially complex routes of the waveguides\cite{Charles2012}. One way this can be achieved is using three-dimensional waveguides in direct-write chips made via ULI (Section \ref{sec_uli}), designed to be stable and resistant to differential strain and temperature effects. The pupil of the telescope is reimaged onto a microlens array mounted to the end face of the 3-dimensional photonic chip, which injects each subaperture into a corresponding waveguide. An example of this method is found in the Dragonfly\cite{Jovanovic2012} and subsequently GLINT instrument\cite{Norris2020b, Martinod2021}, detailed in Section \ref{sec_nulling}, where waveguides are designed as three-dimensional Bezier curves, optimised numerically for optical path length, waveguide separation, bend radius, and minimal length. Alternatively, a set of individual single-mode fibres can be used, with the pupil reimaged into a two-dimensional fibre array via a set of microlenses. This is the approach taken by the FIRST pupil remapping interferometer\cite{Huby2012, Vievard2020}, which formats the outputs of the fibres into a linear non-redundant array forming spectrally-dispersed fringes on the detector, producing extremely accurate interferometric observations\cite{Vievard2023b}. However a major challenge with this method is keeping the separate fibres perfectly path-length-matched to nanometre levels, even in the face of temperature swings and vibrations of an observatory. Currently external free-space optic delay lines can be used, but active on-chip phase modulation is the next step, as described in Section \ref{sec_interf-future}.

\subsection{Beam combination}
But the biggest impact to this domain has been in the interferometric beam combination itself. Rather than creating free space fringes on a detector, integrated optical devices take light from the individual telescopes or sub-apertures and perform the necessary splitting, differential delay, coupling and dispersion to produce a set of discrete outputs encoding the required interferometric information. The highest profile such instrument is the GRAVITY instrument\cite{GRAVITY2017} at the Very Large Telescope Interferometer (VLTI), which combines the light from four 8~m telescopes with the longest baseline being 130~m. This has produced a slew of high impact science results, including the direct measurement of the black hole at the centre of our galaxy via close orbiting stars, directly confirming Einstein's theory of general relativity\cite{GRAVITY2018}. 

\begin{figure}[h]
\centering\includegraphics[width=1\textwidth]{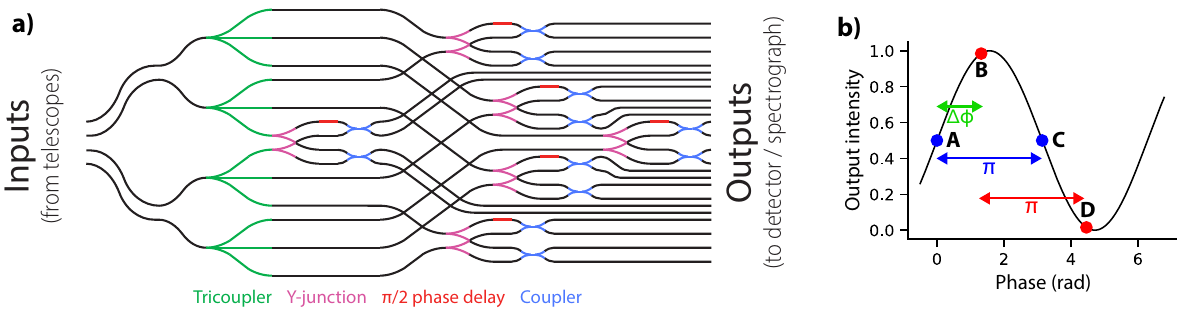}
\caption{An example of a photonic integrated circuit used to perform ABCD beam combination for astronomical interferometry, based on the design published by Benisty, et al.\cite{Benisty2009}. To perform extreme angular-resolution imaging, light from multiple (in this case 4) telescopes or sub-apertures is combined and interfered pair-wise to recover baseline amplitudes and phases. But instead of forming fringes on a detector, the ABCD beam combiner uses an array of photonic components (a) to perform discrete points of interference, effectively quadrature-encoding the virtual interference `fringe' in the fluxes of 4 output waveguides (A,B, C \& D) for each baseline (b). See text for details.}
\label{fig_abcd}
\end{figure}

Another notable implementation is the upcoming beam combiner chip in the FIRST v2 instrument\cite{Lallement2023}. This instrument employs a beam combiner chip using the so called ABCD design\cite{Benisty2009}, as shown in Figure \ref{fig_abcd}. Here, each baseline's effective `fringes' are encoded through four intensity outputs, which are then directly imaged, often with spectral dispersion. This process begins with each telescope or sub-pupil being injected into an input waveguide of a chip. From there, the input beam is split by y-junctions or (in the case of the chip in Figure \ref{fig_abcd}) tricouplers, ensuring each input can be paired and combined with every other input. These y-junctions and couplers function through evanescent coupling between closely positioned waveguides, with the split-ratio being determined by the separation and length of the regions where the waveguides are in proximity. As each waveguide pair reaches the combining section, they are further divided into two by a y-junction. In these junctions a 50/50 intensity split ensures that the two outputs are $\pi$~rad out of phase. Then these four waveguides are combined as two pairs in a directional coupler. But one of these four arms incorporates an (ideally achromatic) $\sim\pi/2$ phase delay (denoted by $\Delta\phi$ in the figure), achieved by varying both the width and the length of the delay region to finely tune dispersion characteristics. This arrangement ultimately yields four outputs per baseline, each offset by $\sim\pi/2$~rad, encoding the fringe. For enhanced accuracy, the delay at the inputs is often rapidly scanned. This temporal modulation contributes to extra calibration precision, providing a more accurate representation of the fringe.

Another major advantage of photonic beam combiners (especially in pupil-remapping applications) is they allow a portion of the injected light from each telescope to be split off to a photometric channel -- this allows the degeneracy between spatial coherence function and differential telescope amplitude shown in Equation \ref{eqn_vis2coh} to be resolved, removing a major source of error.

\subsection{Nulling and habitable exoplanet imaging}
\label{sec_nulling}
While precise measurement of interferometric quantities can produce super-diffraction-limited imaging through the atmosphere, there is one major limitation which no degree of measurement accuracy can overcome -- that of stellar photon noise. 

The most demanding case is that of exoplanet imaging, where a distant planet needs to be resolved separately from its host star. Directly imaging exoplanets is key not just to unambiguously determining their population, orbit, mass, etc., but it allows the spectrum of the light from the planet to be obtained. This is key to characterisation of their atmospheric composition, surface structure, weather and biosignatures. While planets widely separated from their host stars have been imaged using adaptive optics and coronagraphic techniques, these planets reside in freezing, dark orbits similar to that of Pluto around our own sun, or beyond. However, astronomers want to image planets at Earth-like orbits, in the so-called habitable zone, the region where water can remain liquid. Even for the nearest stars, the separation between star and habitable-zone planet is of order 10~milliarcseconds, right at the diffraction limit of the largest telescopes.

While these spatial scales can be imaged by interferometry, the problem is the planet (visible via the light it reflects from its star) is far, far fainter than the host star, with contrast ratios of $\sim10^{-5}$ for close-in Jupiter-sized planets in the most favourable circumstances through to $\sim10^{-10}$ for rocky Earth-like planets. In a standard interferometric measurement the light from both star and planet are mixed and the signal extracted through analysis of the interferometric observables, but this means the Poissonian photon noise from the star is superimposed on the faint planet signal. 
Thus the only hope of measuring the planet signal is to remove the starlight \emph{before} it is measured.

Nulling interferometry involves introducing phase delays between telescopes or sub-apertures to create maximal destructive interference of the central star's light, ideally resulting in a completely dark null when observing an unresolved source, and with any deviation from the null representing the presence of an extended object. But the ability to create the null is highly sensitive to chromaticity, optical path difference and amplitude matching, polarisation effects, seeing, crosstalk, and so forth (collectively known as instrumental leakage). 

Since its initial proposition\cite{Bracewell1978} various forms of nulling interferometry have been proposed, ranging from multiple baseline re-combinations\cite{Angel1997} to multi-element space-based instruments\cite{Leger1996}, with standard formalisms developed\cite{Serabyn2000}. Nulling interferometers like the Keck Interferometer Nuller\cite{Colavita2009} and the Large Binocular Telescope Interferometer\cite{Defrere2016} implement nulling interferometry with conventional bulk optics. But a fundamental limitation of this implementation is the differential spatial wavefront structure of each telescope or sub-aperture induced by seeing, preventing the complete cancellation of light.

To overcome these limitations photonic technologies have been employed, utilising either single-mode fibres, as seen in the Palomar Fibre Nuller\cite{Mennesson2011, Kuhn2015} and the Vortex Fibre Nuller\cite{Ruane2018,Echeverri2023}, or more intricate waveguide systems inscribed within photonic chips, like in the GLINT nuller\cite{Norris2020b, Martinod2021, Lagadec2021, Norris2023}. The single-mode waveguides eliminate higher-order spatial structures, with the nulling effect determined solely by a single phase and amplitude value for each input, based on specific wavelength and polarisation. Advantages offered by photonic chips for nulling include their ability to realise  complex designs incorporating multiple beam combiners and splitters, enabling various simultaneous outputs that encode photometry, bright (constructive interference) channels, and other metrology features, and the ability to include on-chip modulation of phase and amplitude to optimise null depth (see Section \ref{sec_onchipmod}).

The GLINT photonic nuller\cite{Norris2020b, Martinod2021, Lagadec2021, Norris2023}, installed at the Subaru Telescope on the SCExAO\cite{Jovanovic2015, Lozi2018} extreme adaptive optics system, uses a single-mode photonic chip produced by ULI (Section \ref{sec_uli}) to sample the telescope pupil, split light for photometry and beam combination, and perform nulling interference via directional couplers. The telescope pupil is reimaged onto a microlens array (of 30~$\mu$m pitch) attached to the end face of the chip, with each microlens focusing a sub-region of the pupil into a corresponding waveguide. At an intermediate pupil plane a segmented MEMS deformable mirror, with each segment matched to a given microlens, is used to optimise injection via tip/tilt adjustment and scanned in piston to control phase delay. The original GLINT instrument used 4 subapertures and demonstrated the resolved measurement of extended structure (the edge of a star) at spatial resolutions up to 4 times finer than the formal telescope diffraction limit\cite{Martinod2021}, including measuring (via the analysis of measured null depth) the angular diameter of the star $\delta$~Vir to be 10.9$\pm$0.1 milliarcseconds, despite the telescope diffraction limit (8~m aperture at $\lambda$=1.6$\mu$m) being $\sim$40 milliarcseconds, and the seeing-limited resolution being $\sim$500 milliarcseconds. 

But the sensitivity and achievable null-depth of that instrument was low due to the small fraction of the telescope pupil utilised and the residual seeing-induced phase errors across each baseline. Thus an upgraded instrument is now being built, with some key innovations. Rather than directional couplers, the beam combination will be performed in the chip using tricouplers\cite{Martinod2021b, KlinnerTeo2022, Martinod2022}. In addition to the nulled output, these also provide two bright channels whose differential intensity encodes the phase across the baseline -- this can then be used in a closed loop control system to stabilise the null depth by active phase modulation. Furthermore these couplers are inherently achromatic, exploiting symmetry to create a null across a broad bandwidth. The tricoupler is comprised of three single-mode waveguides with overlapping mode fields in a defined interaction region. This can be implemented by starting with the waveguides widely separated, then bringing them together for the required length and separation, and then having them, diverge again. For nuller applications, two of the input waveguides are illuminated, similar to a pair-wise interferometer. The three  waveguides can be arranged in a plane or -- as in the case of the GLINT ULI device -- in a 3D equilateral triangle. In both cases, the tricoupler offers a key advantage to nulling instruments: if the left and right input channels receive anti-phase beams, the structure's symmetry prevents light from entering the central guide. This central waveguide serves as the null channel, and its achromatic, robust null is ensured by the symmetric design, regardless of slight coupling variations in the actual device.
The creation of this $\pi$ radian phase difference between its two inputs requires an achromatic phase delay, which is also being developed as part of this project. Other advancements include greater use of telescope pupil area, reduced stray light and other features - see Section \ref{sec_uli} for more details. Future developments also improve null-depth accuracy by employing a self-calibrating observable -- kernel nulling\cite{Cvetojevic2022} -- analogous to the closure phase used in interferometry.

\subsection{Future innovations for photonic interferometry}
\label{sec_interf-future}

\subsubsection{Active on-chip modulation}
\label{sec_onchipmod}
In all these cases it is critical that the light arriving at the beam combiner from the separate telescopes or sub-apertures is coherent, i.e. the optical path length traversed by the separate beam is perfectly matched. This is hard enough in the case of pupil remapping interferometry, where light incident upon the different fibres is in phase and all that needs to be corrected for is the differential optical fibre length and subsequent drift with temperature, vibration, etc. But in the case of long baseline interferometry much larger optical path length differences need to be controlled -- the multiple telescopes are all situated on the ground but the distance between each telescope and the distant star varies as the earth rotates, requiring active delay lines capable of accurately modulating the differential optical path length between the beams by hundreds of metres. This is an extreme engineering challenge and is a large contributor to the cost of long baseline interferometers. 

An exciting future prospect currently seeing significant research effort is replacing these free-space beam combiners with on-chip phase delay capabilities. One possible implementation would be modulating the waveguide effective refractive index via heating from micro-heaters built directly onto the chip surface\cite{Ji2019}, and novel implementations of this specifically for astronomical interferometry have been explored\cite{Cheriton2022}. Work on developing these for ULI-based astrophotonic chips is underway, with key performance capabilities yet to be constrained. These requirements include the speed of modulation (ideally sub-millisecond time scales to keep up with atmospheric seeing), cross-talk between a heater and adjacent waveguides, maximum delay obtainable, and bandwidth. Another method is to build chips using a nonlinear material such as LiNbO$_3$ and exploiting the electro-optic effect to rapidly modulate the waveguide effective index via electrodes deposited onto the chip\cite{Martin2014, Lallement2023}. But such chips currently face challenges with throughput, as discussed earlier. 

In the case of long baseline interferometry where hundreds of metres of delay may be needed, a hybrid solution may be realised. The large but slow delays (corresponding mostly to sky rotation) may be controlled via single-mode-fibre delay lines such as fibre stretchers or switchers, and the fast atmosphere- and vibration-induced delays implemented on chip. An exciting offshoot of this is that, since light must be injected into SM waveguides anyway, the injection could take place at the telescopes and light transmitted via single mode fibre to the beam combining lab, which would be more versatile and economical than the vacuum beam lines currently employed.

These requirements become far more stringent in the case of nulling interferometry. Here, the phase of the incident beams must be matched to within just a few nanometres in order to create destructive interference (nulls) dark enough to reveal the faint planet light being sought (frequently needing contrasts better than $\sim10^{-6}$ ). Such precise phase control can likely only be achieved with active delay lines within a photonic chip being driven in closed loop by phase-sensing outputs of the chip, in order to avoid non-common-path between phase sensor and delay line. But it is not just the phase that must be matched -- in order to obtain deep nulls (equivalent to high fringe visibility), the \emph{amplitude} of the combined beams must also be perfectly matched (see Equation \ref{eqn_vis2coh}). The actual amplitude of the injected light will in reality vary significantly, due to atmospheric effects (such as scintillation) but also to the high sensitivity of single-mode injection efficiency to the rapidly varying wavefront error. To correct for this, sets of active delay lines and couplers could be implemented in a Mach-Zehnder configuration, allowing the full complex amplitude of each input waveguide to be controlled. 

To further increase this challenge, it is crucial (especially in the nulling application) that the induced phase modulation (and any resulting amplitude modulation) is entirely achromatic. That is, any chosen phase shift of $\phi$~radians applied produces a $\phi$~radian shift for every wavelength across the entire observing bandpass (which for the astronomical H band would be from $1.5\mu m$ to $1.8\mu m$). This requirement applies to both static and active on-chip delays, and precludes the use of a simple delay line, with more complicated devices being required. 

In addition to interferometry, another important application for on-chip modulation would be to combine this with a photonic lantern (Section \ref{sec_pl}) to create an entirely on-chip adaptive optics system for optimal SMF injection. The photonic lantern's single-mode outputs would be used for wavefront sensing information, which is then used to drive on-chip phase and amplitude modulation to allow all outputs to be coherently combined into one single-mode output, instead of applying correction via an external deformable mirror. Wavefront sensing could be implemented by including on the chip sets of couplers  (similar to the ABCD beam combiner in Figure \ref{fig_abcd}) which encode differential input phase into chip output intensities, such that maximum coherent combination corresponds to minima in these outputs. Then the control loop servos on these to minimise their intensity, thus maximising the output of the coherently combined outputs. Alternatively, a generic set of couplers are used and the chip output is split via a dichroic mirror, with wavelengths outside the optimised science band used for tracking. Or, a time domain modulation could be used as previously demonstrated in coherent on-chip combination\cite{Zhang2021}.
This would massively reduce the cost and complexity of a system, ideal for arrays of small telescopes feeding single-mode spectrographs or interferometers. A simpler, hybrid solution could also be realised, wherein the existing AO system is relied upon to deliver a high Strehl-ratio beam, and (slow) on-chip delays used to compensate for internal undesired delays in the chip.

On the more distant horizon, and beyond the scope of this current paper, is the concept of quantum optical interferometry\cite{Nomerotski2020,Ellis2024}. Here, techniques drawn from applied quantum physics fields (especially quantum computing and communications) could be used to remove the requirement for a physical, real-time optical connection (such as beam-lines or optical fibres) between telescopes in an interferometer, for example by using quantum networks\cite{Khabiboulline2019}.

\begin{figure}[t]
\centering\includegraphics[width=0.7\textwidth]{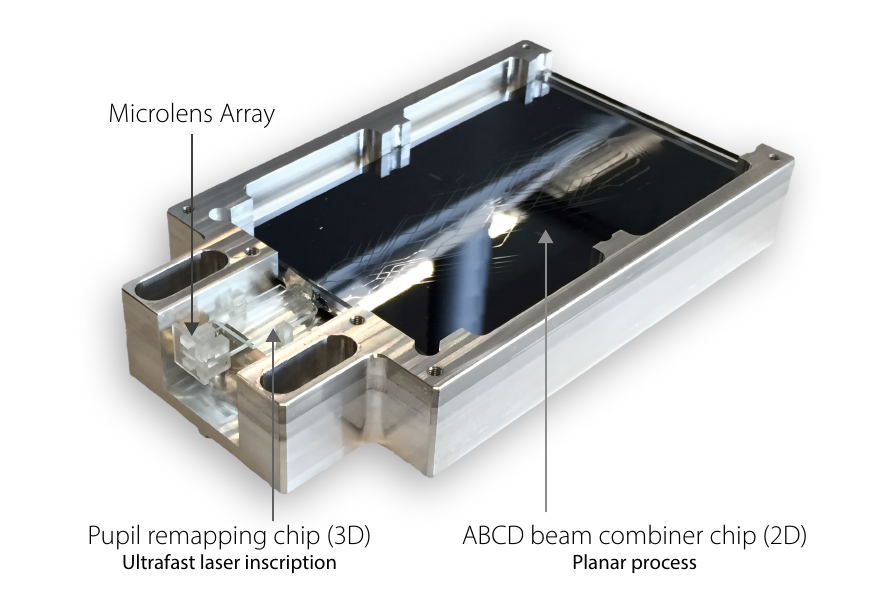}
\caption{An example of a hybrid photonic device, which combines an ultrafast-laser-inscription (ULI) component with a planar component, bonded together and butt-coupled via waveguides of matched mode field diameter and separation. This leverages the 3-dimensional capabilities of ULI (required to coherently sample and remap a two dimensional telescope pupil into a linear waveguide array) and photonic circuit complexity of a planar device.}
\label{fig_hybridchip}
\end{figure}

\subsubsection{Advanced photonic chip fabrication}
\label{sec_chipfab}
{\bf Planar technologies}\\
Integrated optical devices used for these applications are primarily fabricated using two technologies: lithography, and direct-write (also known as ultrafast laser inscription (ULI)). Each technology has distinct advantages and capabilities in producing chips with the required waveguides, splitters, and couplers. It is also possible to have a `best of both worlds' approach with the use of a hybrid device. Such a prototype device has been demonstrated consisting of a 3-dimensional pupil-remapping section made using ULI bonded to a planar chip to perform beam combination \cite{Cvetojevic2021} - see Figure \ref{fig_hybridchip}.
Lithography involves a complex production process with numerous steps, akin to the manufacturing of electronic integrated circuits. 
The process begins with the deposition of films onto a substrate, for example SiO$_2$ onto a Si substrate. Then a sensitised resist coating is applied, which will allow desired regions to be either retained or removed in subsequent etching steps. In photolithography, a two-dimensional waveguide design is transcribed onto an optical mask. This design is then projected onto the layer of photoresist. Alternatively (especially when very fine features or rapid iteration are desired) e-beam lithography can be used, where the pattern is written directly onto the resist using a scanning electron beam. 
The subsequent step involves etching away (using chemical or ion etching) or chemically altering specific regions to create the desired waveguide structure, with the entire process sometimes repeated multiple times to produce the final device\cite{Malbet1999}. Post-processing steps such as applying protective passivisation layers, dicing, and packaging are then performed.

This method is adaptable for various waveguide designs and substrates, such as silica-on-silicon for visible and near-infrared applications, and materials like chalcogenide, LiNbO$_3$, and ZBLAN for mid-infrared wavelengths\cite{Labeye2006}. 
Initially driven by telecommunications requirements for the 1.3~$\mu$m to 1.5~$\mu$m range (matching the astronomical H band), these near-IR (and visible) chips are based on silicon and its oxides, such as silica-on-silicon\cite{Takahashi2003}, silicon-on-insulator\cite{Siew2021} and silicon nitride\cite{Xiang2022}. Silica-on-silicon technology has been extended to operate at wavelengths as long as 2.45~$\mu$m by the GRAVITY instrument, albeit with somewhat low transmission at this wavelength ($\sim$23\%) \cite{GRAVITY2017}.
But extension to the mid-IR requires different materials, and the adaption of existing fabrication techniques to these new platforms. Astrophotonics-focused mid-IR devices, including MMI couplers destined for nulling applications, have been developed using chalcogenide glass\cite{KenchingtonGoldsmith2017} for operation in the L band (3.7~$\mu$m to 4.2~$\mu$m), with waveguides losses of around 0.2~dB/cm.
Silicon and silicon-germanium photonics holds promise for operation at mid-IR wavelengths potentially as far out as 20~$\mu$m\cite{Lin2018, Ren2019}, as do III-V semiconductors such as GaAs\cite{Haas2019}, but have not yet seen development for astrophotonic applications.

A defining characteristic of photolithographic devices is their two-dimensional nature; in general no waveguide can cross above or below one another (although novel multi-layer technologies are being developed\cite{Sacher2017}), and the design of components like couplers and splitters is confined to two dimensions. While the production process can be lengthy due to multiple steps and potentially long time requirements for mask production, photolithography offers significant benefits. For example, it enables the use of materials such as SiN which offer high refractive-index contrast, allowing for tight bend radii and the integration of complex circuits with numerous devices onto a single chip. Alternatively, platforms such as silica-on-silicon have index contrasts and mode-field-diameters similar to standard single-mode-fibres, allowing easy high-efficiency coupling without the addition of tapers. In either case, low waveguide losses (<0.1~dB/cm \cite{Kawachi1990,Xiang2022}) are typical.
While waveguide losses are small, the overall throughput of a complete integrated optic chip is more complicated. Additional losses occur from each individual photonic component and waveguide bends amongst other factors, and can rapidly accumulate. For example the silica-on-silicon H-band chip used in the PIONEER beam combiner had a total throughput of $\sim$65\% \cite{Benisty2009}. Additionally, while high index-contrast platforms such as SiN allow for tight bends with low loss, their small mode fields make direct coupling to standard single mode fibres highly inefficient, and some sort of mode-matching taper must be used. This technique has currently achieved $\sim$60\% coupling efficiency of a standard SMF-28 fibre into a 1$\mu$m core LiN waveguide \cite{Cvetojevic2022}.
Perhaps most significantly, lithographic integrated optics are a mature technology with a well-developed ecosystem, and tried-and-tested designs for the various on-chip components exist.\\

\noindent {\bf Laser direct-write}\\
\label{sec_uli}
The other technology increasingly being utilised is laser direct write, a.k.a. ULI. In these devices full three-dimensional structures can be created, which is especially useful for pupil-remapping applications (as the 2 dimensional telescope pupil can be imaged onto one face of the chip, containing a 2 dimensional array of waveguides, each sampling the desired sub-aperture). 

\begin{figure}[h]
\centering\includegraphics[width=0.8\textwidth]{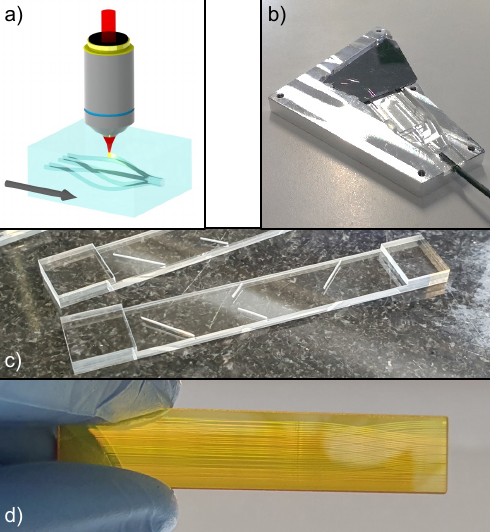}
\caption{a) Basic laser inscription setup showing an ultrashort-pulsed laser focused into a transparent substrate while the substrate is being translated transversely to the stationary laser focus. b) Packaged hybrid assembly of an integrated optics spectrograph (arrayed waveguide grating) coupled to an integrated photonic lantern and multimode optical fibre. c) Integrated optics nulling beam combiner with stray-light suppression slots for the GLINT instrument. d) Mid-infrared (L'-band) 4-telescope beam combiners in a chalcogenide glass sample.}
\label{fig_uli}
\end{figure}

The origins of ULI date back to the seminal papers by Davis et al.\cite{Davis1996} and Glezer et al.\cite{Glezer1996}. In short, an ultrashort-pulsed laser with a pulse duration less than 400 fs and commonly a wavelength of 515, 800 or 1030~nm is focused into a transparent dielectric while the substrate is translated with respect to the stationary focus (see Figure \ref{fig_uli}a). Due to the substrate's transparency no linear absorption takes place, yet the high peak intensity due to tight focusing and ultrashort pulse duration triggers nonlinear photoionisation localised to the focal spot. The deposited energy results in a structural modification of the substrate material that can manifest itself in a smooth change in the refractive index (positive or negative in sign), self-assembled nanostructures or void formation and optical damage at high peak intensities. The smooth refractive index change is most desirable for creating low loss waveguides. Ultrafast Laser Inscription is attractive for developing novel astronomical instrumentation since it offers rapid prototyping\cite{Thomson2009}, avoiding the costly need for photo masks; the ability to create 3-dimensional waveguide circuits free of waveguide crossings that could  lead to loss and cross-talk; its flexibility in choice of substrate material that is transparent at the desired operation wavelength range (visible, near-infrared and mid-infrared), and the modest refractive index contrast of the waveguides facilitating low-loss coupling from an astronomical telescope of standard optical fibre. Moreover, waveguides and photonic components such as directional couplers can be made free of birefringence, exhibiting identical properties for all polarisation states, through thermal annealing\cite{Arriola2013}, stress-tuning \cite{Fernandes2012, Corrielli2018} or clever adaption of their geometry \cite{Pitsios2017}. 

The flexibility of ULI also extends to the type of waveguide, enabling single-mode waveguides, few-mode as well as multimode waveguides to be fabricated through an adaption in inscription parameters, such as laser power, speed at which the sample is translated through the focus and focusing conditions as well as the geometrical cross section of the waveguide itself \cite{Jovanovic2012b}. Since ULI unlike lithography does not rely on a substrate with a guiding layer for fixed thickness, these waveguides can be created in one and the same substrate to, for instance, create photonic lanterns for efficient transformation of multimode light to many single-mode channels \cite{Spaleniak2013, Thomson2011,Leon-Saval2013}. Such ultrafast laser inscribed photonic lanterns can, for instance, be used in a hybrid assembly to feed arrayed waveguide gratings (AWG) for spectroscopy \cite{Cvetojevic2017} (see Figure \ref{fig_uli}b), yet care has to be taken to avoid modal noise induced by the interface between multimode fibre and integrated photonic lantern. It is also possible to use ULI to integrate the photonic lantern and arrayed waveguide gratings into one and the same substrate in a single fabrication step \cite{Douglass2018}. 

Single-mode waveguides act as vital spatial filters for astronomical interferometry. This has been exploited for Dragonfly, an instrument based on pupil remapping interferometry \cite{Jovanovic2012}. The pupil remapping section features a 3-dimensional waveguide chip with 8 waveguides to sample a telescope's pupil. The 8 waveguides are optically pathlength-matched \cite{Charles2012} and take a side-step to suppress interference with residual stray light \cite{Norris2014}. This instrument was upgraded to become GLINT and permanently installed as sub-module of SCExAO \cite{Jovanovic2016b} to perform single-baseline nulling interferometry with simultaneous photometry \cite{Norris2020b} and later to the first multi-baseline nulling interferometer \cite{Martinod2021}. Further upgrades of the photonics inside GLINT are in progress to enable an achromatic null across the astronomical H-band and the capability for simultaneous wavefront sensing using tricouplers (3x3 couplers) \cite{Martinod2021b,KlinnerTeo2022}. A photo of the next generation beam combiner is shown in Figure \ref{fig_uli}c.

Through careful selection of an appropriate substrate, for instance chalcogenide glass, ultrafast laser inscribed waveguides can be extended to the mid-infrared wavelength region with ease. Low loss waveguides and the basic building blocks, such as Y-junctions and directional couplers, for a mid-infrared beam combiner have been demonstrated in Gallium Lanthanum Sulphide (GLS) glass \cite{Gretzinger2019}. This platform is being investigated as 4-telescope beam combiner for the ASGARD/NOTT instrument \cite{Laugier2023, Defrere2022} based on a double Bracewell architecture utilising achromatic directional couplers and Y-junctions design for the L'-band\cite{Sanny2022}. A photo of a beam combiner prototype is shown in Figure \ref{fig_uli}d.

While the current focus is on terrestrial applications for ultrafast laser inscribed photonic circuits, it is worth noting that preliminary tests have shown that ultrafast laser inscribed waveguides are highly robust to temperature extremes as well as radiation, enabling potential satellite applications \cite{Jovanovic2011, Piacentini2021}.

\subsubsection{Improved sensing and integrated spectroscopy}
\label{sec_interf-sensing-spect}
All current photonic devices used for interferometry culminate in a set of single-mode fibres that are sent to a separate detector or spectrograph. This works well but imposes limitations on the allowable complexity of the photonic circuitry (since all outputs must be routed to the chip end face) and the intervening fibres and connector introduce losses. Ideally, the photo-detection and necessary spectral dispersion would be integrated on the same chip as the photonic interferometry device. Such spectral dispersion may already be achievable using photonic devices such as array waveguide gratings as discussed in Section \ref{sec_spectro}. The dream of integrating sensors on-chip is beginning to look possible thanks to rapid advances being made in the domain of silicon photonics\cite{Margalit2021}. 

On-chip integration aside, there is a pressing need for improved detectors for photonic imaging and interferometry applications. These techniques all rely on directly sampling the seeing-induced wavefront, and hence must take measurements at rates comparable to (or ideally faster than) the atmospheric time, which is $\sim$1~ms. This necessitates not only fast but also very low read noise detectors. These two requirements heavily favour minimising the total number of pixels used, but this limits the spectral resolution that can be employed. In the case of nulling interferometry, extremely high dynamic ranges are required. E.g. for a high contrast exoplanet imaging case, the photometric and wavefront-sensing channels of the chip may have $10^6$ or more times the flux of the nulled channels, which contain the faint exoplanet light. 

One exciting solution to this is the use of MKIDS detector\cite{Mazin2012}. These superconducting detectors have truly zero read noise, and count the arrival of individual photons at microsecond time resolutions. This enables not only high speed measurement but also an arbitrarily high dynamic range (as long as flux does not exceed the system's total count-rate capacity). Furthermore, MKIDS detectors measure the energy of each arriving photon, thus providing low-spectral-resolution wavelength dimensions without the need for any cross-dispersion elements; single-mode fibres could be coupled directly to individual MKIDS pixels, or even have MKIDS and photonics built into the same chip. Alternatively, very high spectral resolutions (desired for exoplanet atmosphere characterisation, for example) could be achieved by implementing an MKIDS detector into an Echelle spectrograph, but using the energy-resolving power of the MKIDS array to separate the Echelle orders rather than a separate cross-dispersing element. Spectrographs based on this method, such as the KIDSpec instrument concept, are in development\cite{OBrien2014, OBrien2020}. Plans are currently underway to feed the SM outputs of the GLINT interferometer to the MEC\cite{Walter2020} MKIDS camera at Subaru Telescope to test this approach.

\section{Photonic spectroscopy}
\label{sec_spectro}

Spectroscopy is at the core of a vast number of astronomical science cases, including the measurement of the chemical makeup of distant stars and planets, the kinematic behaviour of both nearby stars and distant galaxies, the detection of exoplanets, the behaviour of black holes at the heart of distant galaxies, among many other areas of investigation. Astrophotonic technologies are now having a major impact on astronomical spectroscopy, unlocking new levels of stability, sensitivity, efficiency, replicability and resolution and enabling entirely new types of spatially resolved spectroscopy. 

\begin{figure}[h]
\centering\includegraphics[width=0.95\textwidth]{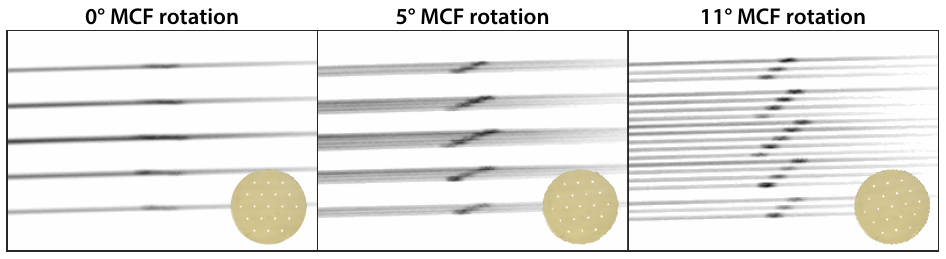}
\caption{Three detector images from a spectrograph using the `tiger' configuration, where the end of a multicore fibre (MCF), such as the output of a photonic lantern, is placed at the slit position instead of the usual linear array. The three images show the results when the hexagonal-patterned MCF is rotated to different angles; it can be seen the optimal angle for this MCF is 11~degrees. Insets show microscope images of the MCF, rotated to the given angles.}
\label{fig_tiger}
\end{figure}

\subsection{Single-mode spectroscopy}
As described in Section \ref{sec_sminj}, it is near impossible to efficiently inject the seeing-aberrated PSF from a telescope directly into a single-mode fibre, due to its extended spatial structure arising from atmospheric turbulence. Conventionally, light is thus fed from telescopes to optical spectrographs via a large core multimode fibre, as such a high mode-count fibre is able to accept seeing-aberrated light. But using multimode fibres leads to significant issues. First is the problem of \emph{modal noise}\cite{Baudrand2001}: the seeing-aberrated PSF injected and environmental changes to the fibre over time means that the mode field at the fibre output is non-uniform -- it is a speckle pattern, and can vary over time. When imaged onto the detector via a diffraction grating or prism, this pattern is mixed with the spectral dispersion, and an ambiguity in fine spectral features is created, a particular problem for precision radial-velocity applications. Then there is a fundamental scaling problem: in order to maintain efficiency, the size of a spectrograph needs to scale with the size of the telescope\cite{Robertson2012}, and (for a seeing limited spectrograph) the cost of the instrument scales as the square of the telescope diameter\cite{Bland-Hawthorn2006}. For current large telescopes this imposes extreme engineering and stability requirements (with high resolution spectrographs being room-sized instruments) and becomes even more problematic for the upcoming Extremely Large Telescopes.

But if light is injected into a single-mode fibre and used to feed a spectrograph, this scaling relation is broken, and very compact but extremely high spectral resolution, stable spectrographs can be made for any telescope size\cite{Bland-Hawthorn2010}. The cost, stability and modal-noise-free nature of such a single-mode spectrograph has now led to the development of extreme precision radial velocity (EPRV) spectrographs, such as the \emph{iLocator}\cite{Crepp2016} for the Large Binocular Telescope, \emph{HISPEC}\cite{Mawet2022} for the Keck telescope and \emph{MODHIS}\cite{Mawet2022} for the coming Thirty Metre Telescope. 

This, of course, presumes the ability to efficiently inject starlight into single-mode fibres in the first place. The aforementioned instruments will use extreme adaptive optics systems to reduce wavefront error and perform efficient injection -- this approach has now achieved over 50\% injection efficiency in H band (1.6~$\mu$m)\cite{Jovanovic2017b} and over 35\% in Y (1.02~$\mu$m) and J (1.22~$\mu$m) bands\cite{Crass2021}. Wavefront correction rapidly becomes more challenging at shorter wavelengths, a particular problem for EPRV studies which are performed at visible wavelengths. But looking to the future, the performance of AO systems at short wavelengths continues to improve as higher actuator-count deformable mirrors come online, and photonic techniques such as the hybrid mode selective photonic lantern\cite{Norris2022c} discussed in Section \ref{sec_sminj} may be used to inject all light into a single waveguide with very high efficiency.

But an economical, passive approach also exists - using a standard photonic lantern (Section \ref{sec_pl}) to take the seeing-aberrated PSF and efficiently convert it into a series of single-mode fibres. Each of these individual single-mode fibres is entirely free of modal noise and can then be placed along a pseudo-slit of a single-mode spectrograph, with the entire signal from the telescope recovered by combining these individual spectra. This concept of a PL-fed single-mode spectrograph has now been extensively studied\cite{Bland-Hawthorn2010,Schwab2012,Betters2013}, and prototype instruments have been designed and successfully demonstrated in the laboratory\cite{Leon-Saval2012,Betters2016}. If the photonic lantern is made using a multicore fibre as its output, the so-called `tiger' configuration can be used, which offers optimal detector usage and instrument simplicity\cite{Betters2014}. Here, instead of  linear array of fibres in a pseudo slit, the multicore-fibre is placed at the spectrograph slit position, and rotated precisely such that the resulting spectra do not overlap -- see Figure \ref{fig_tiger}. This results in a projected fibre separation (around 50~$\mu$m for the example in Figure \ref{fig_tiger}) than is easily achievable with a v-groove fibre array. It also has the benefit that, since adjacent spectra are offset in the wavelength direction, fringing due to coherent mixing from the wings of the adjacent spectrum's PSF is prevented.

\subsection{Fibre Bragg Gratings}
Having the starlight in single-mode fibres also allows other spectral processing operations to be performed before the spectrum is measured. A key example of this is OH suppression. The ability to conduct deep spectroscopy in near-IR (J and H bands) is limited by contamination from the bright emission of excited hydroxyl (OH) radicals in the Earth's atmosphere\cite{Meinel1950}. When undergoing vibrational decay they emit a dense forest of emission lines, and their brightness and multiplicity are such that they cannot be easily subtracted from a measured science spectrum. This is exacerbated by the fact that the relative strengths of the lines are time varying, so cannot be easily calibrated. 

\begin{figure}[h]
\centering\includegraphics[width=0.90\textwidth]{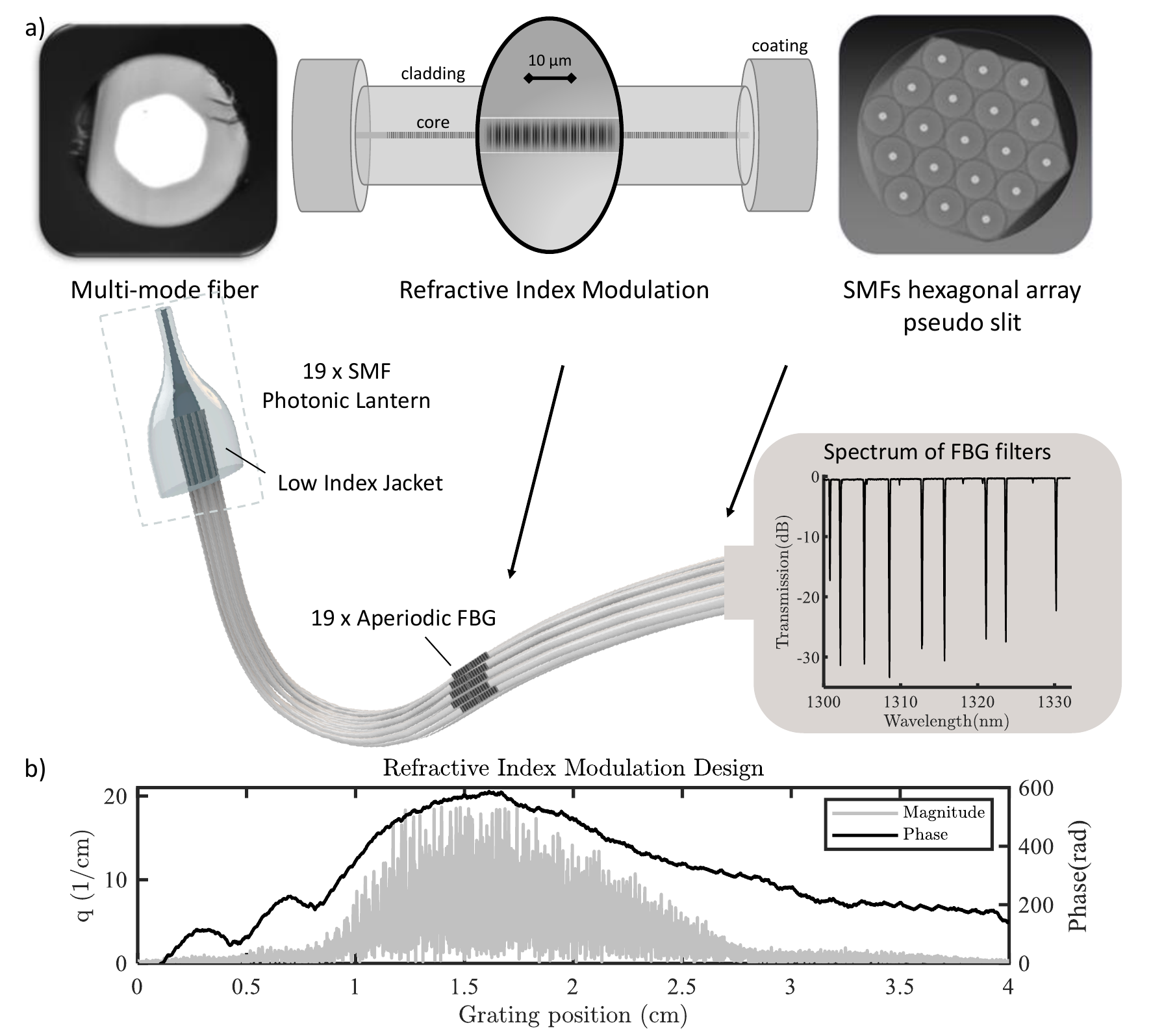}
\caption{Conceptual diagram of an OH suppression unit consisting of a 19-port photonic lantern and 19 aperiodic FBGs in SMFs. In panel a), microscopic views of the multimode end of the lantern and the hexagonal array of SMFs are displayed. In the middle, a zoomed view of the diagram presents the refractive index modulation corresponding to a J band aperiodic FBG. A spectrum of a J band aperiodic FBG is shown on the right. In panel b), the magnitude and phase of the corresponding FBG index modulation are displayed.}
\label{OH_FBG}
\end{figure}

A proposed solution was to filter out the OH lines before they entered the spectrograph, by using a set of fibre Bragg gratings (FBGs) designed to precisely remove the OH lines\cite{Ellis2008}. This has led to the development of on-sky demonstration instruments -- GNOSIS\cite{Trinh2013, Ellis2012} and PRAXIS\cite{Ellis2020} -- which have demonstrated the efficacy of this approach, successfully using FBGs to suppress 103 OH doublets from 1.47$\mu$m to 1.7$\mu$m by factors of up to 40~dB. Since FBGs require single-mode inputs, these instruments used photonic lanterns to efficiently separate the seeing-limited light into an array of single-mode fibres, each connected to its own FBG (see Figure \ref{OH_FBG}). In fact, this FBG-based OH suppression application was one of the main original motivations for the development of the photonic lantern. Going forward, other single-mode filtering devices may be used, such as sets of ring resonators\cite{Ellis2017}, which have the advantage of being integrated onto a photonic chip.

Following the successful demonstration of GNOSIS and PRAXIS, subsequent developments have focused on the multicore fibre Bragg grating\cite{lindley2014demonstration} and J band FBGs\cite{yu2024complete}. In multicore fibre Bragg gratings, each fibre core is leveraged as a distinct spatial channel, within which identical OH suppression FBGs are inscribed to enhance integration. Initial efforts encompassed UV laser inscription within 7-core fibers\cite{lindley2014demonstration} and inscription within 120-core fibers\cite{Birks2012}.

J band FBGs have emerged as the first successfully fabricated OH suppression FBGs within non-telecommunication spectral bands. One of the FBGs demonstrated effective suppression of approximately 9 OH lines or doublets spanning 1300 to 1330 nm, while another suppressed 29 OH lines or doublets within 1270 to 1352 nm\cite{yu2024complete}. These FBGs were inscribed in cladding mode suppressed SMFs produced by Coractive. 

A proposed application of the J band FBG is for the detection of the celestial Positronium recombination $\alpha$ line\cite{ellis2015possibility, ellis2018astrophysical}. Existing Positronium observations are realized based on Gamma-ray emission, and the current best angular resolution of 2.7 degrees is insufficient to resolve the point source distribution of the Positronium. Detection of the Positronium recombination $\alpha$ line at around 1312.22 nm could enable observations with arcsecond-level resolution in the J band.  For this purpose, a design of an OH suppressed integrated photonic spectrograph utilizing J band FBGs has been proposed, which will hopefully be applied to the detection of the Positronium recombination $\alpha$ line\cite{robertson2021seeking}. Currently, it is in the early construction stage\cite{yu2023aperiodic}.

The FBG is fabricated by inscribing a predetermined refractive index modulation into the SMF fibre core using a 244 nm UV laser. The index modulation can be derived from a design procedure tailored for OH suppression FBGs, which is based on the layer-peeling algorithm\cite{yu2024complete}. Since the index modulation design is usually limited by the fabrication constraints in the maximum reachable index modulation and the length, an optimization method based on a genetic algorithm and neural networks can be utilized for precise adjustments in the position and the dephase angle of each partial grating channel\cite{yu2024inverse}.

The inscription apparatus is mainly composed of a phase mask and a Sagnac interferometer to generate a UV interference pattern. The SMFs are Ge-doped and pre-hydrogenated to enhance their photosensitivity so that the UV exposure can elevate the refractive index by an order of magnitude ranging from $\sim10^{-4}$ to $10^{-3}$. The 244 nm UV pattern induces a periodic modulation of the refractive index within the fibre core. In the Sagnac loop, a pair of acoustic-optic modulators are configured to separately detune the phase of each beam. This enables a dithering feature to apply both apodization and chirp functions into the index modulation, and eventually inscribe a sophisticated index modulation design for an aperiodically spaced multi-notch filter\cite{edvell2014optical}.

\begin{figure}[b]
\centering\includegraphics[width=0.80\textwidth]{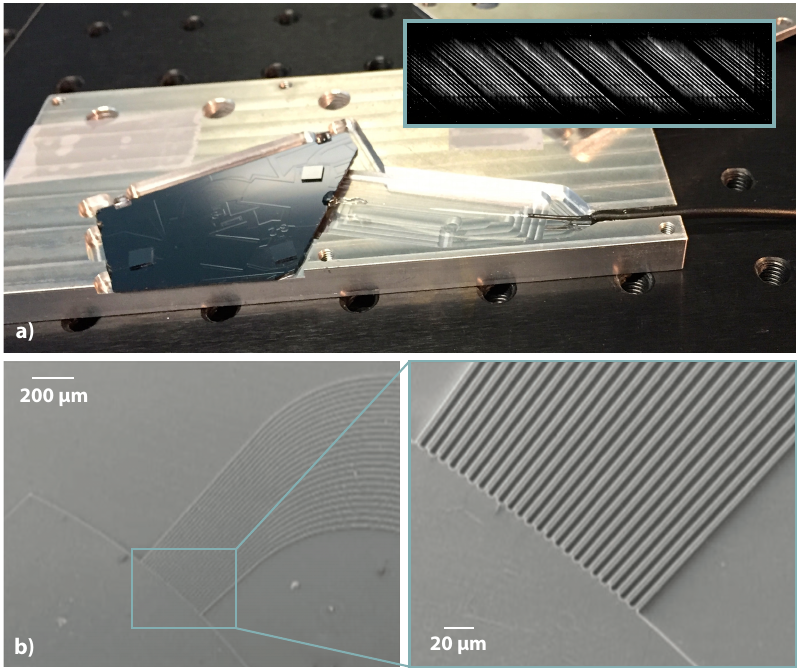}
\caption{a) An arrayed waveguide grating (AWG) photonic spectrograph, packaged for testing. Incident starlight is transmitted via the fibres on the right, and the output spectra emerges from the exposed end face at the bottom of the figure, to be imaged onto a detector. The spectra consist of multiple overlapping orders, which are separated via cross-dispersion to produce a spectrum on the detector as shown in the inset. b) Microscope images of the AWG showing a portion of the waveguide array and its interface with the free propagation region. }
\label{fig_awg}
\end{figure}

\subsection{Integrated photonic spectroscopy}
Beyond using single-mode fibres with conventional optics to produce a diffraction-limited spectrograph, there is great interest in performing the entire spectrographic process within a photonic device. Building an entire SMF-fed spectrograph on a chip -- replacing the usual assembly of collimating lens, diffraction grating or prism and focusing lens -- would yield a number of significant benefits\cite{Bland-Hawthorn2006}. The extreme compactness of a photonic chip and inherent replicability of photonic devices means that a large number of spectrographs could be built in a tiny area, ideal for multi-object or integral-field-unit spectroscopy. For example major galaxy surveys such as Hector\cite{Bryant2020} and the DESI\cite{DESI2016} project could be rapidly scaled, with important impacts for cosmology and galaxy formation science. 
Their small, solid state construction leads to increased thermo-mechanical stability making them ideal for precision radial-velocity applications, as well as for space applications. High spectral-resolution characterisation of exoplanet atmospheres -- using multiple spatial pixels -- is also well suited. The ease and low cost of their replicability makes them a good fit for large spectrographic transient surveys using many small, cheap telescopes. They can also benefit from the active tuning available on photonic devices, and would be ideal for integrating into the same chip as interferometric beam combiners and nullers (Section \ref{sec_interf-future}).

A leading technology for such devices is the arrayed waveguide grating (AWG)\cite{Leijtens2006} (Figure \ref{fig_awg}. In an AWG, light travels from the input single-mode fibre(s) to an input free propagation region (FPR), analogous to a collimating lens. This light then illuminates an array of waveguides, akin to a grating in a traditional spectrograph, which introduces a constant path difference between adjacent waveguides based on the spectral order. Finally, the light from these waveguides is focused in the output FPR, with different wavelengths constructively interfering at distinct spatial locations along the focal plane, thus creating a spectrum.

Significant advances have been made in evaluating and optimising these devices for astronomy\cite{Bland-Hawthorn2006, Gatkine2017,Stoll2021,Gatkine:21}, including modification of existing AWG designs produced for telecommunications\cite{Cvetojevic2012}. Architectures including the use of photonic lanterns to accept seeing-limited telescope light and convert it into a series of single-mode fibres which are then placed along the AWG's input have also been proposed\cite{Bland-Hawthorn2010}.

A modified commercial AWG was used for the first on-sky demonstration of an AWG spectrograph\cite{Cvetojevic2009}, wherein sky background OH lines were measured. Subsequently, extreme adaptive optics were used to efficiently inject light from the 8~m Subaru Telescope into a SMF-fed AWG spectrograph, successfully measuring hydrogen absorption lines in the stellar spectra from a number of stars\cite{Jovanovic2017b}. An integrated photonic lantern fed device was also tested, which led to the discovery of a modal noise issue arising from a mode mismatch between the input MMF and the photonic lantern\cite{Cvetojevic2017} (this can be avoided by instead placing the PL directly at the focal plane).  

However, the demands of astronomy differ from the demands of telecommunications, and thus the throughput, spectral resolution and bandwidth of these demonstration instruments need to be improved upon. Astronomy-optimised AWGs are now being developed, with experimental devices now demonstrating spectral resolutions of >30~000 and efficiencies of 70\% having been demonstrated\cite{Stoll2021}. Future refinements may involve cascading two or more AWGs on a single chip to enhance bandwidth and for order separation\cite{vanWijk2020}, investigation of high index-contrast materials such as silicon nitride\cite{Gatkine2017} and refined designs to minimise loss within the chip.  Related functionality would be to build arbitrary filters on chip, for example to suppress OH lines\cite{Hu2020}. 

While AWGs are currently the technology frontrunner for astrophotonic spectrographs, there are a vast array of other promising techniques\cite{Blind2017} that may prove optimal. For example photonic Fabry-Perot based devices such as cascaded ring resonators\cite{Dai2009} can produce very high spectral resolutions with high bandwidth in an extremely small footprint, and are well suited to detector integration. And stationary-wave integrated Fourier-transform spectrometry (SWIFTS) devices perform Fourier transform spectroscopy on fibre-fed light and directly project the spectral data in two dimensions onto an attached detector\cite{LeCoarer2007}. Looking further ahead, integration of other light processing tasks (such as interferometric beam combination) as well as spectrographic components onto the same chip would yield advantages in stability, replicability and measurement precision, as would integration of novel noise-free detection elements such as MKIDS, as is discussed in Section \ref{sec_interf-sensing-spect}.

\subsection{Spatially resolved spectroscopy with photonic integral field units}
In traditional astronomical spectroscopy, a single spectrum was produced for a single astronomical object at a time, usually by focusing the object of interest onto a spectrograph's slit. Over time spectroscopy has expanded to produce spectra of multiple objects in the field of view at once (multi object spectroscopy) and also to produce spatially resolved spectroscopy where each spatial pixel has a full spectrum (integral field spectroscopy). For high spectral resolution cases, fibre based integral field units are often used. These consist of a 2 dimensional array of fibres at the focal plane, each forming one spatial pixel, and each routed to a separate position along a spectrograph slit to produce its own high resolution spectrum.

\begin{figure}[h]
\centering\includegraphics[width=1.0\textwidth]{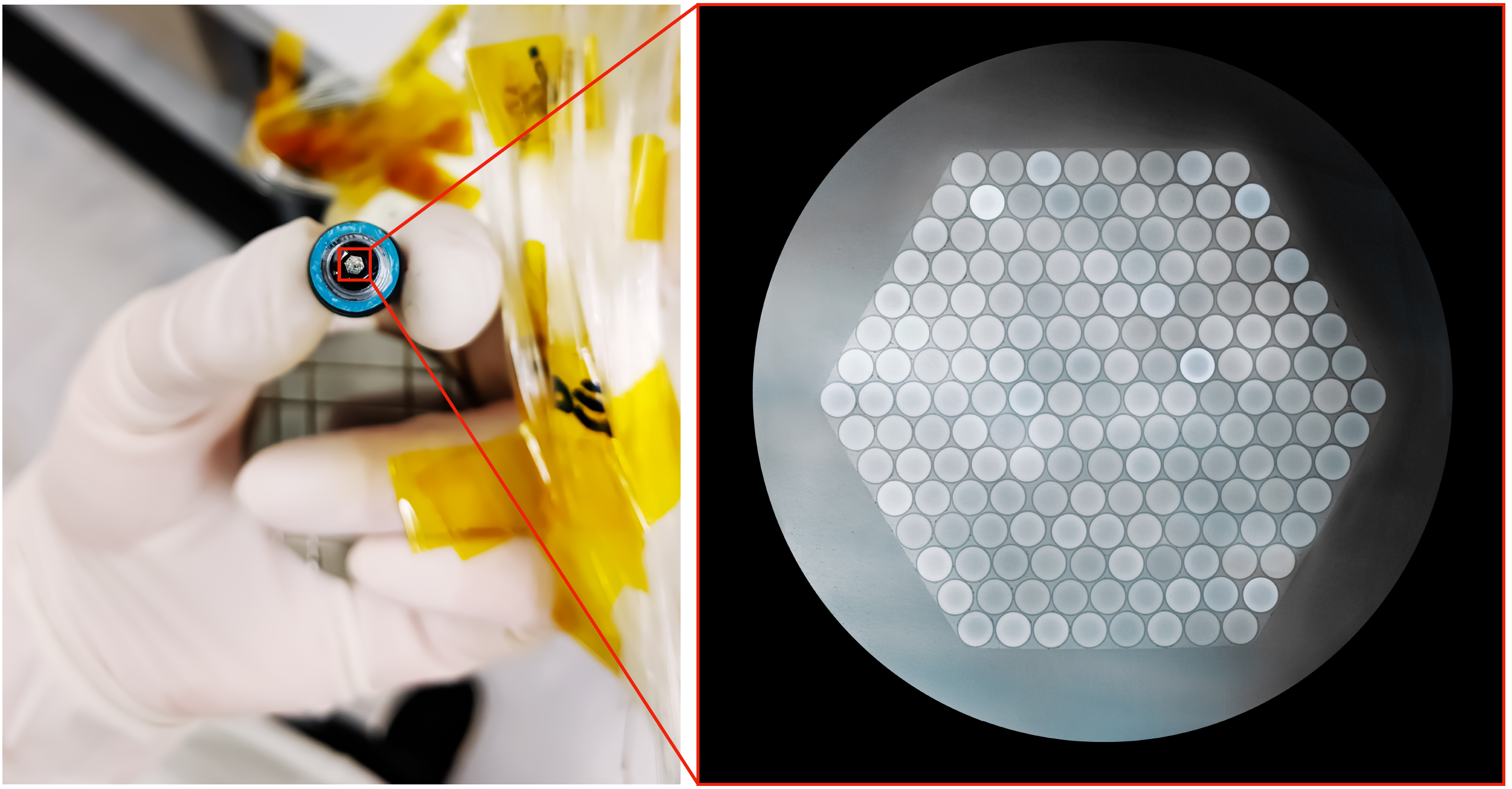}
\caption{Left: a 169 core hexabundle within its ferrule, held in a gloved hand. Right: microscope image of the end of the same hexabundle, with the 169 closely packed cores clearly visible. Each core acts as a spatial pixel, each producing a high resolution spectrum. Microscope image post-processed to remove dust and microscope tiling artefacts.}
\label{fig_hexa}
\end{figure}

Recent photonic developments have now combined these two techniques to produce multi-object, spatially resolved spectroscopy. Multiple different objects over a wide science field of view can be observed at once, and each object is imaged with an array of multiple spatial pixels each having its own high resolution spectrum. One way this can be achieved with photonic integral field units -- tightly packed arrays of optical fibres known as hexabundles\cite{Bland-Hawthorn2011b}, with hexabundles positioned at the locations of the astronomical objects of interest in the telescope focal plane. An image of a hexabundle from the Hector instrument is shown in Figure \ref{fig_hexa}. 
The positioning methods draw on previous single-pixel, multi object spectrographs such as the 2dF instrument\cite{Lewis2002} which used a robotic positioner to place 400 individual fibres at the locations of galaxies over the 56~cm diameter telescope focal plane to an accuracy of 11~$\mu$m, corresponding to a 2 degree field of view. A hexabundle based instrument replaces these individual fibres with a very compact fibre array, having an extremely high fill-factor, more than 50\% higher than any other fibre bundles used in astronomy, to minimise the loss of light between fibre cores. 

The first such instrument to use the hexabundle method was the Sydney-AAO Multi-object Integral field spectrograph (SAMI)\cite{Croom2012,Bryant2015}, at the Anglo-Australian Telescope (AAT). This used 13 hexabundles, each consisting of 61 lightly-fused multimode fibres arranged in a circular packing to provide the spatial resolution. Positioning was performed by plugging each hexabundle into a specially machined field plate with holes to accept each hexabundle at the precise location of the target galaxies. The MaNGA instrument\cite{Bundy2015} utilised a similar positioning method, using 29 fibre-IFUs each having between 7 and 127 fibres, Subsequently the next generation instrument -- Hector\cite{Bryant2020} -- was commissioned, which uses a robotic positioner to accurately deploy 21 hexabundles over a 2 degree field of view, each hexabundle having between 37 and 169 fibres, and is now underway performing the Hector Galaxy Survey at the AAT.

Design and manufacture of hexabundles involved developing new photonic techniques to achieve the astronomical requirements\cite{Bryant2011}. Key requirements were a maximum fill-factor, minimal cross-talk, and high throughput. The initial production of hexabundles included two main types: heavily fused and lightly fused fibre bundles. Firstly the fibre cladding is etched back to a precisely determined thickness, with 2~$\mu$m cladding thickness for 100 $\mu$m core diameter fibres initially found optimal. Heavily fused hexabundles are created by inserting fibres into a glass tube and fusing them at high power, resulting in the elimination of interstitial holes, and non-circular cores. The end of the bundle is inserted into a connector and polished. Lightly fused hexabundles involve a process where the etched fibres are lightly fused, maintaining a balance between minimal cladding, core deformation, and mechanical strength. In initial tests, it was found that, compared to lightly fused devices, heavily fused hexabundles exhibited higher cross-talk, lower throughput and very high focal ratio degradation, due to the compression of the cores. 

Minimising focal ratio degradation (FRD) was the major design driver\cite{Bryant2014} to avoid light loss, and was considered in the optimisation of all design parameters and manufacturing techniques. FRD occurs due to mode-mixing as the light travels down the fibre, corresponding to a non-conservation of etendue in geometric optics terms. This leads to the numerical aperture of the output beam being wider than the input beam, reducing the encircled energy of the resulting PSF and ultimately spectrograph performance. 

The FRD introduced during hexabundle production is affected by the fibre fusing method, fibre end cleaving, hexabundle end surface polish, fibre bending diameter, external pressure on the fibres, splicing, and the recoating process. Thanks to the optimisation of these parameters, the latest generation of hexabundles, as used in the Hector instrument, now introduce no additional FRD beyond that of the original fibre\cite{Bryant2014, Wang2023}. 

Based on the success of hexabundles at optical wavelengths, the future of hexabundle technology lies in extending this capability to infrared wavelengths from $0.9-2.2\,\mu$m. Additionally new devices are incorporating microlenses to increase the application of hexabundles beyond the low f/\# scenarios of wide-field telescopes, to the high f/\# that would feed fibres directly on the larger telescopes. The combination of infrared fibre bundles, with FBGs and built in microlens assemblies will enable the 3D galaxy science currently achieved for $z<0.12$ to be extended to $z\sim0.5$ on large ground-based telescopes.  Gas kinematics, excitation tracers, gas feeding and feedback and the growth of mass and angular momentum traced through star formation would all then be mapped to galaxy environment and large scale structure beyond the local Universe.

\section{Conclusion}
This paper presents a brief overview of the current state and future prospects of astrophotonics, an emerging field at the intersection of astronomy and photonics. Advancements in astrophotonics have the potential to revolutionise observational astronomy by enhancing the capabilities of astronomical instruments in terms of sensitivity, resolution, and versatility.

A key focus is the advancements in photonic wavefront sensing, imaging and interferometry, highlighting the importance of coherent light manipulation and the challenges in both pupil-remapping and long baseline interferometry. Devices such as photonic lanterns are able to sense, at the telescope focal plane, the highly destructive wavefront error caused by atmospheric turbulence in order to help correct it, and can potentially produce super-diffraction-limited imaging from terrestrial telescopes. The development of on-chip achromatic phase delay capabilities including active delay lines, devices which operate over broad bandwidths, and more complex photonic capabilities, are promising steps towards increasing the precision and efficiency of these systems. Central to this and other astrophotonic applications are the advancements in photonic chip fabrication technologies, such as photolithography and ultrafast laser inscription (ULI). This technology has been instrumental in creating complex, three-dimensional waveguide circuits, which are essential for various astronomical applications.

Another key area is the integration of spectroscopic functions onto photonic chips. This development is significant as it offers a path towards more compact and efficient spectrographic systems, including integrating photo-detection and spectral dispersion components on-chip, and breaking the dependence between telescope size and spectrograph size and cost. The role of single-mode fibres in reducing modal noise and enhancing the precision of spectroscopic observations is also discussed. The ongoing development of photonic integral field units (IFUs) and arrayed waveguide gratings (AWGs), among other technologies, will likely be crucial in achieving high-resolution, spatially resolved spectroscopy.

Astrophotonics represents a transformative approach in astronomical instrumentation, standing to offer unprecedented precision and functionality, and its potential advantages will only continue to increase as the next generation of Extremely Large Telescopes comes online. The nexus between photonics and astronomy opens up new possibilities for exploring the universe, and ongoing research and development in this field are expected to yield groundbreaking discoveries in the years to come. The advancements discussed in this paper are just the beginning, and the full potential of astrophotonics is yet to be realised.

\begin{backmatter}

\bmsection{Acknowledgements}
\noindent 
B.Norris is the recipient of an Australian Research Council Discovery Early Career Award (DE210100953) funded by the Australian Government. 

\bmsection{Disclosures}
\noindent The authors declare no conflicts of interest.


\end{backmatter}


\bibliography{bnorrisbib_2023}

\begin{thebibliography}{100}
\newcommand{\enquote}[1]{``#1''}

\bibitem{GRAVITY2018}
{GRAVITY Collaboration}, R.~{Abuter}, A.~{Amorim}, \emph{et~al.}, \enquote{{Detection of the gravitational redshift in the orbit of the star S2 near the Galactic centre massive black hole},} {\protect\JournalTitle{\aap}} \textbf{615}, L15 (2018).

\bibitem{Jovanovic2023}
N.~Jovanovic, P.~Gatkine, N.~Anugu, \emph{et~al.}, \enquote{2023 astrophotonics roadmap: pathways to realizing multi-functional integrated astrophotonic instruments,} {\protect\JournalTitle{Journal of Physics: Photonics}} \textbf{5}, 042501 (2023).

\bibitem{Johns2012}
M.~{Johns}, P.~{McCarthy}, K.~{Raybould}, \emph{et~al.}, \enquote{{Giant Magellan Telescope: overview},} in \emph{Ground-based and Airborne Telescopes IV,}  vol. 8444 of \emph{Society of Photo-Optical Instrumentation Engineers (SPIE) Conference Series} L.~M. {Stepp}, R.~{Gilmozzi}, and H.~J. {Hall}, eds. (2012), p. 84441H.

\bibitem{Sanders2013}
G.~H. {Sanders}, \enquote{{The Thirty Meter Telescope (TMT): An International Observatory},} {\protect\JournalTitle{Journal of Astrophysics and Astronomy}} \textbf{34}, 81--86 (2013).

\bibitem{Gilmozzi2007}
R.~{Gilmozzi} and J.~{Spyromilio}, \enquote{{The European Extremely Large Telescope (E-ELT)},} {\protect\JournalTitle{The Messenger}} \textbf{127}, 11 (2007).

\bibitem{Tamai2022}
R.~{Tamai}, B.~{Koehler}, M.~{Cirasuolo}, \emph{et~al.}, \enquote{{Status of the ESO's ELT construction},} in \emph{Ground-based and Airborne Telescopes IX,}  vol. 12182 of \emph{Society of Photo-Optical Instrumentation Engineers (SPIE) Conference Series} H.~K. {Marshall}, J.~{Spyromilio}, and T.~{Usuda}, eds. (2022), p. 121821A.

\bibitem{Guyon2005}
O.~{Guyon}, \enquote{{Limits of Adaptive Optics for High-Contrast Imaging},} {\protect\JournalTitle{\apj}} \textbf{629}, 592--614 (2005).

\bibitem{Huby2012}
E.~{Huby}, G.~{Perrin}, F.~{Marchis}, \emph{et~al.}, \enquote{{FIRST, a fibered aperture masking instrument. I. First on-sky test results},} {\protect\JournalTitle{Astronomy \& Astrophysics}} \textbf{541}, A55 (2012).

\bibitem{Martin2022}
G.~Martin, M.~Foin, S.~Phatak, \emph{et~al.}, \enquote{{Hybrid electro-optic visible multi-telescope beam combiner for next generation FIRST/SUBARU instruments},} in \emph{Advances in Optical and Mechanical Technologies for Telescopes and Instrumentation V,}  vol. 12188 R.~Navarro and R.~Geyl, eds., International Society for Optics and Photonics (SPIE, 2022), p. 121885Y.

\bibitem{Lallement2023}
M.~{Lallement}, E.~{Huby}, S.~{Lacour}, \emph{et~al.}, \enquote{{Photonic beam-combiner for visible interferometry with Subaru coronagraphic extreme adaptive optics/fibered imager for a single telescope: laboratory characterization and design optimization},} {\protect\JournalTitle{Journal of Astronomical Telescopes, Instruments, and Systems}} \textbf{9}, 025003 (2023).

\bibitem{Norris2014}
B.~{Norris}, N.~{Cvetojevic}, S.~{Gross}, \emph{et~al.}, \enquote{{High-performance 3D waveguide architecture for astronomical pupil-remapping interferometry},} {\protect\JournalTitle{Optics Express}} \textbf{22}, 18335 (2014).

\bibitem{GRAVITY2017}
{GRAVITY Collaboration}, R.~Abuter, M.~Accardo, \emph{et~al.}, \enquote{{First light for GRAVITY: Phase referencing optical interferometry for the Very Large Telescope Interferometer},} {\protect\JournalTitle{Astronomy and Astrophysics}} \textbf{602}, A94 (2017).

\bibitem{GRAVITY2023}
{GRAVITY Collaboration}, F.~{Widmann}, N.~{Schuhler}, \emph{et~al.}, \enquote{{Polarization analysis of the VLTI and GRAVITY},} {\protect\JournalTitle{arXiv e-prints}} arXiv:2311.03472 (2023).

\bibitem{Leon-Saval2005}
S.~G. {Leon-Saval}, T.~A. {Birks}, J.~{Bland-Hawthorn}, and M.~{Englund}, \enquote{{Multimode fiber devices with single-mode performance},} {\protect\JournalTitle{Optics Letters}} \textbf{30}, 2545--2547 (2005).

\bibitem{Leon-Saval2013}
S.~G. {Leon-Saval}, A.~{Argyros}, and J.~{Bland -Hawthorn}, \enquote{{Photonic lanterns},} {\protect\JournalTitle{Nanophotonics}} \textbf{2}, 429--440 (2013).

\bibitem{Birks2015}
T.~A. {Birks}, I.~{Gris-S{\'a}nchez}, S.~{Yerolatsitis}, \emph{et~al.}, \enquote{{The photonic lantern},} {\protect\JournalTitle{Advances in Optics and Photonics}} \textbf{7}, 107 (2015).

\bibitem{Birks2012}
T.~A. {Birks}, B.~J. {Mangan}, A.~{D{\'\i}ez}, \emph{et~al.}, \enquote{{``Photonic lantern'' spectral filters in multi-core Fiber},} {\protect\JournalTitle{Optics Express}} \textbf{20}, 13996 (2012).

\bibitem{Spaleniak2013}
I.~{Spaleniak}, N.~{Jovanovic}, S.~{Gross}, \emph{et~al.}, \enquote{{Integrated photonic building blocks for next-generation astronomical instrumentation II: the multimode to single mode transition},} {\protect\JournalTitle{Optics Express}} \textbf{21}, 27197 (2013).

\bibitem{Norris2020}
B.~R.~M. {Norris}, J.~{Wei}, C.~H. {Betters}, \emph{et~al.}, \enquote{{An all-photonic focal-plane wavefront sensor},} {\protect\JournalTitle{Nature Communications}} \textbf{11}, 5335 (2020).

\bibitem{Gretzinger2019}
T.~Gretzinger, S.~Gross, A.~Arriola, and M.~J. Withford, \enquote{Towards a photonic mid-infrared nulling interferometer in chalcogenide glass,} {\protect\JournalTitle{Opt. Express}} \textbf{27}, 8626--8638 (2019).

\bibitem{Wong2022}
A.~P. {Wong}, B.~R.~M. {Norris}, V.~{Deo}, \emph{et~al.}, \enquote{{Machine learning for wavefront sensing},} in \emph{Adaptive Optics Systems VIII,}  vol. 12185 of \emph{Society of Photo-Optical Instrumentation Engineers (SPIE) Conference Series} L.~{Schreiber}, D.~{Schmidt}, and E.~{Vernet}, eds. (2022), p. 121852I.

\bibitem{Leon-Saval2014}
S.~G. Leon-Saval, N.~K. Fontaine, J.~R. Salazar-Gil, \emph{et~al.}, \enquote{{Mode-selective photonic lanterns for space-division multiplexing},} {\protect\JournalTitle{Optics Express}} \textbf{22}, 1036 (2014).

\bibitem{Velazquez-Benitez2018}
A.~M. Vel{\'a}zquez-Ben{\'\i}tez, J.~E. Antonio-L{\'o}pez, J.~C. Alvarado-Zacar{\'\i}as, \emph{et~al.}, \enquote{Scaling photonic lanterns for space-division multiplexing,} {\protect\JournalTitle{Scientific Reports}} \textbf{8}, 8897 (2018).

\bibitem{Davies2012}
R.~Davies and M.~Kasper, \enquote{{Adaptive Optics for Astronomy},} {\protect\JournalTitle{Annual Review of Astronomy and Astrophysics}} \textbf{50}, 305--351 (2012).

\bibitem{Guyon2018}
O.~Guyon, \enquote{{Extreme Adaptive Optics},} {\protect\JournalTitle{Annual Review of Astronomy and Astrophysics}} \textbf{56}, 315--355 (2018).

\bibitem{Platt2001}
B.~C. Platt and R.~Shack, \enquote{History and principles of shack-hartmann wavefront sensing,} {\protect\JournalTitle{Journal of refractive surgery}} \textbf{17}, S573--S577 (2001).

\bibitem{Ragazzoni1996}
R.~{Ragazzoni}, \enquote{{Pupil plane wavefront sensing with an oscillating prism},} {\protect\JournalTitle{Journal of Modern Optics}} \textbf{43}, 289--293 (1996).

\bibitem{Sauvage2007}
J.-F. {Sauvage}, T.~{Fusco}, G.~{Rousset}, and C.~{Petit}, \enquote{{Calibration and precompensation of noncommon path aberrations for extreme adaptive optics},} {\protect\JournalTitle{Journal of the Optical Society of America A}} \textbf{24}, 2334--2346 (2007).

\bibitem{Sauvage2016}
J.-F. {Sauvage}, T.~{Fusco}, M.~{Lamb}, \emph{et~al.}, \emph{{Tackling down the low wind effect on SPHERE instrument}} (2016), vol. 9909 of \emph{Society of Photo-Optical Instrumentation Engineers (SPIE) Conference Series}, p. 990916.

\bibitem{Milli2018}
J.~{Milli}, M.~{Kasper}, P.~{Bourget}, \emph{et~al.}, \enquote{{Low wind effect on VLT/SPHERE: impact, mitigation strategy, and results},} in \emph{Proceedings of the SPIE,}  vol. 10703 of \emph{Society of Photo-Optical Instrumentation Engineers (SPIE) Conference Series} (2018), p. 107032A.

\bibitem{NDiaye2018}
M.~N'Diaye, F.~Martinache, N.~Jovanovic, \emph{et~al.}, \enquote{{Calibration of the island effect: Experimental validation of closed-loop focal plane wavefront control on Subaru/SCExAO},} {\protect\JournalTitle{Astronomy and Astrophysics}} \textbf{610}, A18 (2018).

\bibitem{Vievard2019}
S.~{Vievard}, S.~{Bos}, F.~{Cassaing}, \emph{et~al.}, \enquote{{Overview of focal plane wavefront sensors to correct for the Low Wind Effect on SUBARU/SCExAO},} {\protect\JournalTitle{arXiv e-prints}} arXiv:1912.10179 (2019).

\bibitem{Mocoeur2009}
I.~Moc{\oe}ur, L.~M. Mugnier, and F.~Cassaing, \enquote{Analytical solution to the phase-diversity problem for real-time wavefront sensing,} {\protect\JournalTitle{Opt. Lett.}} \textbf{34}, 3487--3489 (2009).

\bibitem{Vievard2018}
S.~Vievard, F.~Cassaing, L.~M. Mugnier, \emph{et~al.}, \enquote{{Real-time full alignment and phasing of multiple-aperture imagers using focal-plane sensors on unresolved objects},} {\protect\JournalTitle{SPIE}} \textbf{10698}, 106986F (2018).

\bibitem{Korkiakoski2014}
V.~Korkiakoski, C.~U. Keller, N.~Doelman, \emph{et~al.}, \enquote{{Fast {\&}amp; Furious focal-plane wavefront sensing},} {\protect\JournalTitle{Applied Optics}} \textbf{53}, 4565 (2014).

\bibitem{Martinache2013}
F.~{Martinache}, \enquote{{The Asymmetric Pupil Fourier Wavefront Sensor},} {\protect\JournalTitle{\pasp}} \textbf{125}, 422 (2013).

\bibitem{Martinache2014}
F.~{Martinache}, O.~{Guyon}, N.~{Jovanovic}, \emph{et~al.}, \enquote{{On-Sky Speckle Nulling Demonstration at Small Angular Separation with SCExAO},} {\protect\JournalTitle{\pasp}} \textbf{126}, 565 (2014).

\bibitem{Norris2022b}
B.~R.~M. {Norris}, J.~{Wei}, C.~H. {Betters}, \emph{et~al.}, \enquote{{Demonstration of a photonic-lantern focal-plane wavefront sensor using fiber mode conversion and deep learning},} in \emph{Adaptive Optics Systems VIII,}  vol. 12185 of \emph{Society of Photo-Optical Instrumentation Engineers (SPIE) Conference Series} L.~{Schreiber}, D.~{Schmidt}, and E.~{Vernet}, eds. (2022), p. 1218530.

\bibitem{Lin2021}
J.~{Lin}, N.~{Jovanovic}, and M.~P. {Fitzgerald}, \enquote{{Design considerations of photonic lanterns for diffraction-limited spectrometry},} {\protect\JournalTitle{Journal of the Optical Society of America B Optical Physics}} \textbf{38}, A51 (2021).

\bibitem{Sweeney2021}
D.~{Sweeney}, B.~R.~M. {Norris}, P.~{Tuthill}, \emph{et~al.}, \enquote{{Learning the lantern: neural network applications to broadband photonic lantern modeling},} {\protect\JournalTitle{Journal of Astronomical Telescopes, Instruments, and Systems}} \textbf{7}, 028007 (2021).

\bibitem{Lin2022}
J.~{Lin}, M.~P. {Fitzgerald}, Y.~{Xin}, \emph{et~al.}, \enquote{{Focal-plane wavefront sensing with photonic lanterns: theoretical framework},} {\protect\JournalTitle{Journal of the Optical Society of America B Optical Physics}} \textbf{39}, 2643 (2022).

\bibitem{Lin2022c}
J.~{Lin}, Y.~{Xin}, B.~{Norris}, \emph{et~al.}, \enquote{{Exoplanet detection with photonic lanterns for focal-plane wavefront sensing and control},} in \emph{Adaptive Optics Systems VIII,}  vol. 12185 of \emph{Society of Photo-Optical Instrumentation Engineers (SPIE) Conference Series} L.~{Schreiber}, D.~{Schmidt}, and E.~{Vernet}, eds. (2022), p. 121852G.

\bibitem{Lin2022b}
J.~{Lin}, S.~{Vievard}, N.~{Jovanovic}, \emph{et~al.}, \enquote{{Experimental measurements of AO-fed photonic lantern coupling efficiencies},} in \emph{Society of Photo-Optical Instrumentation Engineers (SPIE) Conference Series,}  vol. 12188 of \emph{Society of Photo-Optical Instrumentation Engineers (SPIE) Conference Series} (2022), p. 121882E.

\bibitem{Xin2022}
Y.~{Xin}, N.~{Jovanovic}, G.~{Ruane}, \emph{et~al.}, \enquote{{Efficient Detection and Characterization of Exoplanets within the Diffraction Limit: Nulling with a Mode-selective Photonic Lantern},} {\protect\JournalTitle{\apj}} \textbf{938}, 140 (2022).

\bibitem{Jovanovic2016}
N.~{Jovanovic}, N.~{Cvetojevic}, C.~{Schwab}, \emph{et~al.}, \enquote{{Efficiently feeding single-mode fiber photonic spectrographs with an extreme adaptive optics system: on-sky characterization and preliminary spectroscopy},} in \emph{Society of Photo-Optical Instrumentation Engineers (SPIE) Conference Series,}  vol. 9908 of \emph{Proceedings of the SPIE} (2016), p. 99080R.

\bibitem{Ellis2021}
S.~C. Ellis, J.~Bland-Hawthorn, and S.~G. Leon-Saval, \enquote{General coupling efficiency for fiber-fed astronomical instruments,} {\protect\JournalTitle{J. Opt. Soc. Am. B}} \textbf{38}, A64--A74 (2021).

\bibitem{Norris2022c}
B.~{Norris}, C.~{Betters}, J.~{Wei}, \emph{et~al.}, \enquote{{Optimal broadband starlight injection into a single-mode fibre with integrated photonic wavefront sensing},} {\protect\JournalTitle{Optics Express}} \textbf{30}, 34908 (2022).

\bibitem{Kim2024}
Y.~J. {Kim}, M.~P. {Fitzgerald}, J.~{Lin}, \emph{et~al.}, \enquote{{Coherent Imaging with Photonic Lanterns},} {\protect\JournalTitle{\apj}} \textbf{964}, 113 (2024).

\bibitem{Norris2024ip}
B.~{Norris} \emph{et~al.} (2024). {in prep}.

\bibitem{Shao1992}
M.~{Shao} and M.~M. {Colavita}, \enquote{{Long-baseline optical and infrared stellar interferometry.}} {\protect\JournalTitle{\araa}} \textbf{30}, 457--498 (1992).

\bibitem{Monnier2003}
J.~D. {Monnier}, \enquote{{Optical interferometry in astronomy},} {\protect\JournalTitle{Reports on Progress in Physics}} \textbf{66}, 789--857 (2003).

\bibitem{Labeyrie2006}
A.~{Labeyrie}, S.~G. Lipson, and P.~Nisenson, \emph{An introduction to Optical Stellar Interferometry} (Cambridge University Press, 2006).

\bibitem{Haguenauer2012}
P.~{Haguenauer}, R.~{Abuter}, L.~{Andolfato}, \emph{et~al.}, \enquote{{The Very Large Telescope Interferometer v2012+},} in \emph{Optical and Infrared Interferometry III,}  vol. 8445 of \emph{Society of Photo-Optical Instrumentation Engineers (SPIE) Conference Series} F.~{Delplancke}, J.~K. {Rajagopal}, and F.~{Malbet}, eds. (2012), p. 84450D.

\bibitem{tenBrummelaar2005}
T.~A. {ten Brummelaar}, H.~A. {McAlister}, S.~T. {Ridgway}, \emph{et~al.}, \enquote{{First Results from the CHARA Array. II. A Description of the Instrument},} {\protect\JournalTitle{\apj}} \textbf{628}, 453--465 (2005).

\bibitem{Tuthill2000}
P.~G. {Tuthill}, J.~D. {Monnier}, W.~C. {Danchi}, \emph{et~al.}, \enquote{{Michelson Interferometry with the Keck I Telescope},} {\protect\JournalTitle{\pasp}} \textbf{112}, 555--565 (2000).

\bibitem{Baldwin1986}
J.~E. Baldwin, C.~A. Haniff, C.~D. Mackay, and P.~J. Warner, \enquote{{Closure phase in high-resolution optical imaging},} {\protect\JournalTitle{Nature (ISSN 0028-0836)}} \textbf{320}, 595 (1986).

\bibitem{duForesto1998}
V.~{Coud{\'e} du Foresto}, G.~{Perrin}, C.~{Ruilier}, \emph{et~al.}, \enquote{{FLUOR fibered instrument at the IOTA interferometer},} in \emph{Astronomical Interferometry,}  vol. 3350 of \emph{Society of Photo-Optical Instrumentation Engineers (SPIE) Conference Series} R.~D. {Reasenberg}, ed. (1998), pp. 856--863.

\bibitem{Anugu2020}
N.~{Anugu}, J.-B. {Le Bouquin}, J.~D. {Monnier}, \emph{et~al.}, \enquote{{MIRC-X: A Highly Sensitive Six-telescope Interferometric Imager at the CHARA Array},} {\protect\JournalTitle{\aj}} \textbf{160}, 158 (2020).

\bibitem{Charles2012}
N.~{Charles}, N.~{Jovanovic}, S.~{Gross}, \emph{et~al.}, \enquote{{Design of optically path-length-matched, three-dimensional photonic circuits comprising uniquely routed waveguides},} {\protect\JournalTitle{\ao}} \textbf{51}, 6489 (2012).

\bibitem{Jovanovic2012}
N.~{Jovanovic}, P.~G. {Tuthill}, B.~{Norris}, \emph{et~al.}, \enquote{{Starlight demonstration of the Dragonfly instrument: an integrated photonic pupil-remapping interferometer for high-contrast imaging},} {\protect\JournalTitle{Monthly Notices of the Royal Astronomical Society}} \textbf{427}, 806--815 (2012).

\bibitem{Norris2020b}
B.~R.~M. {Norris}, N.~{Cvetojevic}, T.~{Lagadec}, \emph{et~al.}, \enquote{{First on-sky demonstration of an integrated-photonic nulling interferometer: the GLINT instrument},} {\protect\JournalTitle{Monthly Notices of the Royal Astronomical Society}} \textbf{491}, 4180--4193 (2020).

\bibitem{Martinod2021}
M.-A. {Martinod}, B.~{Norris}, P.~{Tuthill}, \emph{et~al.}, \enquote{{Scalable photonic-based nulling interferometry with the dispersed multi-baseline GLINT instrument},} {\protect\JournalTitle{Nature Communications}} \textbf{12}, 2465 (2021).

\bibitem{Vievard2020}
S.~{Vievard}, E.~{Huby}, S.~{Lacour}, \emph{et~al.}, \enquote{{FIRST, a pupil-remapping fiber interferometer at the Subaru Telescope: on-sky results},} in \emph{Society of Photo-Optical Instrumentation Engineers (SPIE) Conference Series,}  vol. 11446 of \emph{Society of Photo-Optical Instrumentation Engineers (SPIE) Conference Series} (2020), p. 1144629.

\bibitem{Vievard2023b}
S.~{Vievard}, E.~{Huby}, S.~{Lacour}, \emph{et~al.}, \enquote{{Single-aperture spectro-interferometry in the visible at the Subaru telescope with FIRST: First on-sky demonstration on Keho`oea ({\ensuremath{\alpha}} Lyrae) and Hokulei ({\ensuremath{\alpha}} Aurigae)},} {\protect\JournalTitle{\aap}} \textbf{677}, A84 (2023).

\bibitem{Benisty2009}
M.~{Benisty}, J.-P. {Berger}, L.~{Jocou}, \emph{et~al.}, \enquote{{An integrated optics beam combiner for the second generation VLTI instruments},} {\protect\JournalTitle{Astronomy \& Astrophysics}} \textbf{498}, 601--613 (2009).

\bibitem{Bracewell1978}
R.~N. Bracewell, \enquote{{Detecting nonsolar planets by spinning infrared interferometer},} {\protect\JournalTitle{Nature}} \textbf{274}, 780--781 (1978).

\bibitem{Angel1997}
J.~R.~P. Angel and N.~J. Woolf, \enquote{{An Imaging Nulling Interferometer to Study Extrasolar Planets},} {\protect\JournalTitle{Astrophysical Journal}} \textbf{475}, 373--379 (1997).

\bibitem{Leger1996}
A.~{L{\'e}ger}, J.~M. {Mariotti}, B.~{Mennesson}, \emph{et~al.}, \enquote{{Could We Search for Primitive Life on Extrasolar Planets in the Near Future?}} {\protect\JournalTitle{\icarus}} \textbf{123}, 249--255 (1996).

\bibitem{Serabyn2000}
E.~Serabyn, \enquote{{Nulling interferometry: symmetry requirements and experimental results},} {\protect\JournalTitle{Proc. SPIE Vol. 4006}} \textbf{4006}, 328--339 (2000).

\bibitem{Colavita2009}
M.~M. {Colavita}, E.~{Serabyn}, R.~{Millan-Gabet}, \emph{et~al.}, \enquote{{Keck Interferometer Nuller Data Reduction and On-Sky Performance},} {\protect\JournalTitle{\pasp}} \textbf{121}, 1120 (2009).

\bibitem{Defrere2016}
D.~{Defr{\`e}re}, P.~M. {Hinz}, B.~{Mennesson}, \emph{et~al.}, \enquote{{Nulling Data Reduction and On-sky Performance of the Large Binocular Telescope Interferometer},} {\protect\JournalTitle{\apj}} \textbf{824}, 66 (2016).

\bibitem{Mennesson2011}
B.~Mennesson, C.~Hanot, E.~Serabyn, \emph{et~al.}, \enquote{{HIGH-CONTRAST STELLAR OBSERVATIONS WITHIN THE DIFFRACTION LIMIT AT THE PALOMAR HALE TELESCOPE},} {\protect\JournalTitle{Astrophysical Journal}} \textbf{743}, 178 (2011).

\bibitem{Kuhn2015}
J.~K{\"u}hn, B.~Mennesson, K.~Liewer, \emph{et~al.}, \enquote{{EXPLORING INTERMEDIATE (5-40 AU) SCALES AROUND AB AURIGAE WITH THE PALOMAR FIBER NULLER},} {\protect\JournalTitle{Astrophysical Journal}} \textbf{800}, 55 (2015).

\bibitem{Ruane2018}
G.~Ruane, J.~Wang, D.~Mawet, \emph{et~al.}, \enquote{Efficient spectroscopy of exoplanets at small angular separations with vortex fiber nulling,} {\protect\JournalTitle{The Astrophysical Journal}} \textbf{867}, 143 (2018).

\bibitem{Echeverri2023}
D.~Echeverri, J.~Xuan, N.~Jovanovic, \emph{et~al.}, \enquote{{Vortex fiber nulling for exoplanet observations: implementation and first light},} {\protect\JournalTitle{Journal of Astronomical Telescopes, Instruments, and Systems}} \textbf{9}, 035002 (2023).

\bibitem{Lagadec2021}
T.~{Lagadec}, B.~{Norris}, S.~{Gross}, \emph{et~al.}, \enquote{{The GLINT South testbed for nulling interferometry with photonics: Design and on-sky results at the Anglo-Australian Telescope},} {\protect\JournalTitle{Publications of the Astronomical Society of Australia}} \textbf{38}, e036 (2021).

\bibitem{Norris2023}
B.~R. Norris, M.-A. Martinod, P.~G. Tuthill, \emph{et~al.}, \enquote{{Machine-learning approach for optimal self-calibration and fringe tracking in photonic nulling interferometry},} {\protect\JournalTitle{Journal of Astronomical Telescopes, Instruments, and Systems}} \textbf{9}, 048005 (2023).

\bibitem{Jovanovic2015}
N.~{Jovanovic}, F.~{Martinache}, O.~{Guyon}, \emph{et~al.}, \enquote{{The Subaru Coronagraphic Extreme Adaptive Optics System: Enabling High-Contrast Imaging on Solar-System Scales},} {\protect\JournalTitle{\pasp}} \textbf{127}, 890 (2015).

\bibitem{Lozi2018}
J.~Lozi, O.~Guyon, N.~Jovanovic, \emph{et~al.}, \enquote{{Characterizing vibrations at the Subaru Telescope for the Subaru coronagraphic extreme adaptive optics instrument},} {\protect\JournalTitle{Journal of Astronomical Telescopes, Instruments, and Systems}} \textbf{4}, 1 -- 13 (2018).

\bibitem{Martinod2021b}
M.-A. {Martinod}, P.~{Tuthill}, S.~{Gross}, \emph{et~al.}, \enquote{{Achromatic photonic tricouplers for application in nulling interferometry},} {\protect\JournalTitle{\ao}} \textbf{60}, D100 (2021).

\bibitem{KlinnerTeo2022}
T.~{Klinner-Teo}, M.-A. {Martinod}, P.~{Tuthill}, \emph{et~al.}, \enquote{{Achromatic design of a photonic tricoupler and phase shifter for broadband nulling interferometry},} {\protect\JournalTitle{Journal of Astronomical Telescopes, Instruments, and Systems}} \textbf{8}, 045001 (2022).

\bibitem{Martinod2022}
M.-A. {Martinod}, T.~{Klinner-Teo}, P.~{Tuthill}, \emph{et~al.}, \enquote{{Achromatic nulling interferometry and fringe tracking with 3D-photonic tricouplers with GLINT},} in \emph{Optical and Infrared Interferometry and Imaging VIII,}  vol. 12183 of \emph{Society of Photo-Optical Instrumentation Engineers (SPIE) Conference Series} A.~{M{\'e}rand}, S.~{Sallum}, and J.~{Sanchez-Bermudez}, eds. (2022), p. 121830K.

\bibitem{Cvetojevic2022}
N.~Cvetojevic, F.~Martinache, P.~Chingaipe, \emph{et~al.}, \enquote{3-beam self-calibrated kernel nulling photonic interferometer,} {\protect\JournalTitle{Astronomy and Astrophysics-A\&A}} \textbf{663}, 46 (2022).

\bibitem{Ji2019}
X.~Ji, X.~Yao, Y.~Gan, \emph{et~al.}, \enquote{{On-chip tunable photonic delay line},} {\protect\JournalTitle{APL Photonics}} \textbf{4}, 090803 (2019).

\bibitem{Cheriton2022}
R.~{Cheriton}, S.~{Janz}, G.~{Herriot}, \emph{et~al.}, \enquote{{Integrated astrophotonic phase control for high resolution optical interferometry},} in \emph{Advances in Optical and Mechanical Technologies for Telescopes and Instrumentation,}  vol. 12188 of \emph{Society of Photo-Optical Instrumentation Engineers (SPIE) Conference Series} (2022), p. 121883J.

\bibitem{Martin2014}
G.~{Martin}, S.~{Heidmann}, F.~{Thomas}, \emph{et~al.}, \enquote{{Lithium Niobate active beam combiners: results of on-chip fringe locking, fringe scanning and high contrast integrated optics interferometry and spectrometry},} in \emph{Optical and Infrared Interferometry IV,}  vol. 9146 (2014), p. 91462I.

\bibitem{Zhang2021}
B.~Zhang, J.~Sun, C.~Lao, \emph{et~al.}, \enquote{All-fiber photonic lantern multimode optical receiver with coherent adaptive optics beam combining,}  (2021).

\bibitem{Nomerotski2020}
A.~{Nomerotski}, P.~{Stankus}, A.~{Slosar}, \emph{et~al.}, \enquote{{Quantum-assisted optical interferometers: instrument requirements},} in \emph{Optical and Infrared Interferometry and Imaging VII,}  vol. 11446 of \emph{Society of Photo-Optical Instrumentation Engineers (SPIE) Conference Series} P.~G. {Tuthill}, A.~{M{\'e}rand}, and S.~{Sallum}, eds. (2020), p. 1144617.

\bibitem{Ellis2024}
S.~C. {Ellis} and J.~{Bland-Hawthorn}, \enquote{{Astrophotonics: recent and future developments},} {\protect\JournalTitle{arXiv e-prints}} arXiv:2404.02368 (2024).

\bibitem{Khabiboulline2019}
E.~T. {Khabiboulline}, J.~{Borregaard}, K.~{De Greve}, and M.~D. {Lukin}, \enquote{{Optical Interferometry with Quantum Networks},} {\protect\JournalTitle{Physical Review Letters}} \textbf{123}, 070504 (2019).

\bibitem{Cvetojevic2021}
N.~{Cvetojevic}, B.~R.~M. {Norris}, S.~{Gross}, \emph{et~al.}, \enquote{{Building hybridized 28-baseline pupil-remapping photonic interferometers for future high-resolution imaging},} {\protect\JournalTitle{\ao}} \textbf{60}, D33 (2021).

\bibitem{Malbet1999}
F.~Malbet, P.~Kern, I.~Schanen-Duport, \emph{et~al.}, \enquote{{Integrated optics for astronomical interferometry - I. Concept and astronomical applications},} {\protect\JournalTitle{Astronomy and Astrophysics Supplement Series}} \textbf{138}, 135--145 (1999).

\bibitem{Labeye2006}
P.~Labeye, J.-E. Broquin, J.-P. Berger, \emph{et~al.}, \enquote{{Silicon-based integrated optics for stellar interferometry imaging},} {\protect\JournalTitle{Silicon Photonics. Edited by Kubby}} \textbf{6125}, 161--171 (2006).

\bibitem{Takahashi2003}
H.~Takahashi, \enquote{{Planar lightwave circuit devices for optical communication: present and future},} in \emph{Active and Passive Optical Components for WDM Communications III,}  vol. 5246 A.~K. Dutta, A.~A.~S. Awwal, N.~K. Dutta, and K.~Fujiura, eds., International Society for Optics and Photonics (SPIE, 2003), pp. 520 -- 531.

\bibitem{Siew2021}
S.~Y. Siew, B.~Li, F.~Gao, \emph{et~al.}, \enquote{Review of silicon photonics technology and platform development,} {\protect\JournalTitle{Journal of Lightwave Technology}} \textbf{39}, 4374--4389 (2021).

\bibitem{Xiang2022}
C.~Xiang, W.~Jin, and J.~E. Bowers, \enquote{Silicon nitride passive and active photonic integrated circuits: trends and prospects,} {\protect\JournalTitle{Photon. Res.}} \textbf{10}, A82--A96 (2022).

\bibitem{KenchingtonGoldsmith2017}
H.-D.~K. Goldsmith, M.~Ireland, P.~Ma, \emph{et~al.}, \enquote{Improving the extinction bandwidth of mmi chalcogenide photonic chip based mir nulling interferometers,} {\protect\JournalTitle{Opt. Express}} \textbf{25}, 16813--16824 (2017).

\bibitem{Lin2018}
H.~Lin, Z.~Luo, T.~Gu, \emph{et~al.}, {\protect\JournalTitle{Nanophotonics}} \textbf{7}, 393--420 (2018).

\bibitem{Ren2019}
H.~Ren, L.~Shen, A.~F.~J. Runge, \emph{et~al.}, \enquote{Low-loss silicon core fibre platform for mid-infrared nonlinear photonics,} {\protect\JournalTitle{Light: Science \& Applications}} \textbf{8}, 105 (2019).

\bibitem{Haas2019}
J.~Haas, P.~Artmann, and B.~Mizaikoff, \enquote{Mid-infrared gaas/algaas micro-ring resonators characterized via thermal tuning,} {\protect\JournalTitle{RSC Adv.}} \textbf{9}, 8594--8599 (2019).

\bibitem{Sacher2017}
W.~D. Sacher, J.~C. Mikkelsen, P.~Dumais, \emph{et~al.}, \enquote{Tri-layer silicon nitride-on-silicon photonic platform for ultra-low-loss crossings and interlayer transitions,} {\protect\JournalTitle{Opt. Express}} \textbf{25}, 30862--30875 (2017).

\bibitem{Kawachi1990}
M.~Kawachi, \enquote{Silica waveguides on silicon and their application to integrated-optic components,} {\protect\JournalTitle{Optical and Quantum Electronics}} \textbf{22}, 391--416 (1990).

\bibitem{Davis1996}
K.~M. Davis, K.~Miura, N.~Sugimoto, and K.~Hirao, \enquote{Writing waveguides in glass with a femtosecond laser,} {\protect\JournalTitle{Optics letters}} \textbf{21}, 1729--1731 (1996).

\bibitem{Glezer1996}
E.~Glezer, M.~Milosavljevic, L.~Huang, \emph{et~al.}, \enquote{Three-dimensional optical storage inside transparent materials,} {\protect\JournalTitle{Optics letters}} \textbf{21}, 2023--2025 (1996).

\bibitem{Thomson2009}
R.~R. {Thomson}, A.~K. {Kar}, and J.~{Allington-Smith}, \enquote{{Ultrafast laser inscription: an enabling technology for astrophotonics},} {\protect\JournalTitle{Optics Express}} \textbf{17}, 1963--1969 (2009).

\bibitem{Arriola2013}
A.~{Arriola}, S.~{Gross}, N.~{Jovanovic}, \emph{et~al.}, \enquote{{Low bend loss waveguides enable compact, efficient 3D photonic chips},} {\protect\JournalTitle{Optics Express}} \textbf{21}, 2978 (2013).

\bibitem{Fernandes2012}
L.~A. Fernandes, J.~R. Grenier, P.~R. Herman, \emph{et~al.}, \enquote{Stress induced birefringence tuning in femtosecond laser fabricated waveguides in fused silica,} {\protect\JournalTitle{Optics express}} \textbf{20}, 24103--24114 (2012).

\bibitem{Corrielli2018}
G.~Corrielli, S.~Atzeni, S.~Piacentini, \emph{et~al.}, \enquote{Symmetric polarization-insensitive directional couplers fabricated by femtosecond laser writing,} {\protect\JournalTitle{Optics express}} \textbf{26}, 15101--15109 (2018).

\bibitem{Pitsios2017}
I.~Pitsios, F.~Samara, G.~Corrielli, \emph{et~al.}, \enquote{Geometrically-controlled polarisation processing in femtosecond-laser-written photonic circuits,} {\protect\JournalTitle{Scientific reports}} \textbf{7}, 11342 (2017).

\bibitem{Jovanovic2012b}
N.~Jovanovic, I.~Spaleniak, S.~Gross, \emph{et~al.}, \enquote{Integrated photonic building blocks for next-generation astronomical instrumentation i: the multimode waveguide,} {\protect\JournalTitle{Optics Express}} \textbf{20}, 17029--17043 (2012).

\bibitem{Thomson2011}
R.~{Thomson}, T.~{Birks}, S.~{Leon-Saval}, \emph{et~al.}, \enquote{{Ultrafast laser inscription of an integrated photonic lantern},} {\protect\JournalTitle{Optics Express}} \textbf{19}, 5698--5705 (2011).

\bibitem{Cvetojevic2017}
N.~Cvetojevic, N.~Jovanovic, S.~Gross, \emph{et~al.}, \enquote{Modal noise in an integrated photonic lantern fed diffraction-limited spectrograph,} {\protect\JournalTitle{Opt. Express}} \textbf{25}, 25546--25565 (2017).

\bibitem{Douglass2018}
G.~Douglass, F.~Dreisow, S.~Gross, and M.~Withford, \enquote{Femtosecond laser written arrayed waveguide gratings with integrated photonic lanterns,} {\protect\JournalTitle{Optics express}} \textbf{26}, 1497--1505 (2018).

\bibitem{Jovanovic2016b}
N.~{Jovanovic}, O.~{Guyon}, J.~{Lozi}, \emph{et~al.}, \enquote{{The SCExAO high contrast imager: transitioning from commissioning to science},} in \emph{Society of Photo-Optical Instrumentation Engineers (SPIE) Conference Series,}  vol. 9909 of \emph{Proceedings of the SPIE} (2016), p. 99090W.

\bibitem{Laugier2023}
R.~{Laugier}, D.~{Defr{\`e}re}, J.~{Woillez}, \emph{et~al.}, \enquote{{Asgard/NOTT: L-band nulling interferometry at the VLTI. I. Simulating the expected high-contrast performance},} {\protect\JournalTitle{\aap}} \textbf{671}, A110 (2023).

\bibitem{Defrere2022}
D.~{Defr{\`e}re}, A.~{Bigioli}, C.~{Dandumont}, \emph{et~al.}, \enquote{{L-band nulling interferometry at the VLTI with Asgard/Hi-5: status and plans},} in \emph{Optical and Infrared Interferometry and Imaging VIII,}  vol. 12183 of \emph{Society of Photo-Optical Instrumentation Engineers (SPIE) Conference Series} A.~{M{\'e}rand}, S.~{Sallum}, and J.~{Sanchez-Bermudez}, eds. (2022), p. 121830H.

\bibitem{Sanny2022}
A.~Sanny, S.~Gross, L.~Labadie, \emph{et~al.}, \enquote{Development of the 4-telescope photonic nuller of hi-5 for the characterization of exoplanets in the mid-ir,} in \emph{Optical and Infrared Interferometry and Imaging VIII,}  vol. 12183 (SPIE, 2022), pp. 425--434.

\bibitem{Jovanovic2011}
N.~Jovanovic, S.~Armatys, S.~Gross, \emph{et~al.}, \enquote{Prospects for integrated photonics in space applications,} in \emph{2011 International Quantum Electronics Conference (IQEC) and Conference on Lasers and Electro-Optics (CLEO) Pacific Rim incorporating the Australasian Conference on Optics, Lasers and Spectroscopy and the Australian Conference on Optical Fibre Technology,}  (IEEE, 2011), pp. 1925--1927.

\bibitem{Piacentini2021}
S.~Piacentini, T.~Vogl, G.~Corrielli, \emph{et~al.}, \enquote{Space qualification of ultrafast laser-written integrated waveguide optics,} {\protect\JournalTitle{Laser \& Photonics Reviews}} \textbf{15}, 2000167 (2021).

\bibitem{Margalit2021}
N.~Margalit, C.~Xiang, S.~M. Bowers, \emph{et~al.}, \enquote{{Perspective on the future of silicon photonics and electronics},} {\protect\JournalTitle{Applied Physics Letters}} \textbf{118}, 220501 (2021).

\bibitem{Mazin2012}
B.~A. {Mazin}, B.~{Bumble}, S.~R. {Meeker}, \emph{et~al.}, \enquote{{A superconducting focal plane array for ultraviolet, optical, and near-infrared astrophysics},} {\protect\JournalTitle{Optics Express}} \textbf{20}, 1503 (2012).

\bibitem{OBrien2014}
K.~O'Brien, N.~Thatte, and B.~Mazin, \enquote{{KIDSpec: an MKID based medium resolution integral field spectrograph},} in \emph{Ground-based and Airborne Instrumentation for Astronomy V,}  vol. 9147 S.~K. Ramsay, I.~S. McLean, and H.~Takami, eds., International Society for Optics and Photonics (SPIE, 2014), p. 91470G.

\bibitem{OBrien2020}
K.~O'Brien, \enquote{Kidspec: An mkid-based medium-resolution, integral field spectrograph,} {\protect\JournalTitle{Journal of Low Temperature Physics}} \textbf{199}, 537--546 (2020).

\bibitem{Walter2020}
A.~B. {Walter}, N.~{Fruitwala}, S.~{Steiger}, \emph{et~al.}, \enquote{{The MKID Exoplanet Camera for Subaru SCExAO},} {\protect\JournalTitle{\pasp}} \textbf{132}, 125005 (2020).

\bibitem{Baudrand2001}
J.~{Baudrand} and G.~A.~H. {Walker}, \enquote{{Modal Noise in High-Resolution, Fiber-fed Spectra: A Study and Simple Cure},} {\protect\JournalTitle{\pasp}} \textbf{113}, 851--858 (2001).

\bibitem{Robertson2012}
J.~G. {Robertson} and J.~{Bland-Hawthorn}, \enquote{{Compact high-resolution spectrographs for large and extremely large telescopes: using the diffraction limit},} in \emph{Ground-based and Airborne Instrumentation for Astronomy IV,}  vol. 8446 of \emph{Society of Photo-Optical Instrumentation Engineers (SPIE) Conference Series} I.~S. {McLean}, S.~K. {Ramsay}, and H.~{Takami}, eds. (2012), p. 844623.

\bibitem{Bland-Hawthorn2006}
J.~Bland-Hawthorn and A.~Horton, \enquote{{Instruments without optics: an integrated photonic spectrograph},} in \emph{Ground-based and Airborne Instrumentation for Astronomy,}  vol. 6269 I.~S. McLean and M.~Iye, eds., International Society for Optics and Photonics (SPIE, 2006), p. 62690N.

\bibitem{Bland-Hawthorn2010}
J.~{Bland-Hawthorn}, J.~{Lawrence}, G.~{Robertson}, \emph{et~al.}, \enquote{{PIMMS: photonic integrated multimode microspectrograph},} in \emph{Ground-based and Airborne Instrumentation for Astronomy III,}  vol. 7735 of \emph{Proceedings of the SPIE} (2010), p. 77350N.

\bibitem{Crepp2016}
J.~R. {Crepp}, J.~{Crass}, D.~{King}, \emph{et~al.}, \enquote{{iLocater: a diffraction-limited Doppler spectrometer for the Large Binocular Telescope},} in \emph{Ground-based and Airborne Instrumentation for Astronomy VI,}  vol. 9908 of \emph{Society of Photo-Optical Instrumentation Engineers (SPIE) Conference Series} C.~J. {Evans}, L.~{Simard}, and H.~{Takami}, eds. (2016), p. 990819.

\bibitem{Mawet2022}
D.~Mawet, M.~P. Fitzgerald, Q.~Konopacky, \emph{et~al.}, \enquote{{Fiber-fed high-resolution infrared spectroscopy at the diffraction limit with Keck-HISPEC and TMT-MODHIS: status update},} in \emph{Ground-based and Airborne Instrumentation for Astronomy IX,}  vol. 12184 C.~J. Evans, J.~J. Bryant, and K.~Motohara, eds., International Society for Optics and Photonics (SPIE, 2022), p. 121841R.

\bibitem{Jovanovic2017b}
N.~{Jovanovic}, C.~{Schwab}, O.~{Guyon}, \emph{et~al.}, \enquote{{Efficient injection from large telescopes into single-mode fibres: Enabling the era of ultra-precision astronomy},} {\protect\JournalTitle{Astronomy \& Astrophysics}} \textbf{604}, A122 (2017).

\bibitem{Crass2021}
J.~{Crass}, A.~{Bechter}, B.~{Sands}, \emph{et~al.}, \enquote{{Final design and on-sky testing of the iLocater SX acquisition camera: broad-band single-mode fibre coupling},} {\protect\JournalTitle{\mnras}} \textbf{501}, 2250--2267 (2021).

\bibitem{Schwab2012}
C.~Schwab, S.~G. Leon-Saval, C.~H. Betters, \emph{et~al.}, \enquote{Single mode, extreme precision doppler spectrographs,} {\protect\JournalTitle{Proceedings of the International Astronomical Union}} \textbf{8}, 403--406 (2012).

\bibitem{Betters2013}
C.~H. {Betters}, S.~G. {Leon-Saval}, J.~G. {Robertson}, and J.~{Bland-Hawthorn}, \enquote{{Beating the classical limit: A diffraction-limited spectrograph for an arbitrary input beam},} {\protect\JournalTitle{Optics Express}} \textbf{21}, 26103 (2013).

\bibitem{Leon-Saval2012}
S.~G. {Leon-Saval}, C.~H. {Betters}, and J.~{Bland -Hawthorn}, \emph{{The Photonic TIGER: a multicore fiber-fed spectrograph}} (2012), vol. 8450 of \emph{Society of Photo-Optical Instrumentation Engineers (SPIE) Conference Series}, p. 84501K.

\bibitem{Betters2016}
C.~H. Betters, A.~Murray, J.~Bland-Hawthorn, and S.~G. Leon-Saval, \enquote{{Precision radial velocities with inexpensive compact spectrographs},} in \emph{Ground-based and Airborne Instrumentation for Astronomy VI,}  vol. 9908 C.~J. Evans, L.~Simard, and H.~Takami, eds., International Society for Optics and Photonics (SPIE, 2016), pp. 367 -- 374.

\bibitem{Betters2014}
C.~H. {Betters}, S.~G. {Leon-Saval}, J.~{Bland-Hawthorn}, \emph{et~al.}, \enquote{{PIMMS {\'e}chelle: the next generation of compact diffraction limited spectrographs for arbitrary input beams},} in \emph{Ground-based and Airborne Instrumentation for Astronomy V,}  vol. 9147 of \emph{Proceedings of the SPIE} (2014), p. 91471I.

\bibitem{Meinel1950}
I.~A.~B. {Meinel}, \enquote{{OH Emission Bands in the Spectrum of the Night Sky.}} {\protect\JournalTitle{\apj}} \textbf{111}, 555 (1950).

\bibitem{Ellis2008}
S.~C. {Ellis} and J.~{Bland-Hawthorn}, \enquote{{The case for OH suppression at near-infrared wavelengths},} {\protect\JournalTitle{\mnras}} \textbf{386}, 47--64 (2008).

\bibitem{Trinh2013}
C.~Q. {Trinh}, S.~C. {Ellis}, J.~{Bland-Hawthorn}, \emph{et~al.}, \enquote{{GNOSIS: The First Instrument to Use Fiber Bragg Gratings for OH Suppression},} {\protect\JournalTitle{\aj}} \textbf{145}, 51 (2013).

\bibitem{Ellis2012}
S.~C. {Ellis}, J.~{Bland-Hawthorn}, J.~{Lawrence}, \emph{et~al.}, \enquote{{Suppression of the near-infrared OH night-sky lines with fibre Bragg gratings - first results},} {\protect\JournalTitle{\mnras}} \textbf{425}, 1682--1695 (2012).

\bibitem{Ellis2020}
S.~C. {Ellis}, J.~{Bland-Hawthorn}, J.~S. {Lawrence}, \emph{et~al.}, \enquote{{First demonstration of OH suppression in a high-efficiency near-infrared spectrograph},} {\protect\JournalTitle{\mnras}} \textbf{492}, 2796--2806 (2020).

\bibitem{Ellis2017}
S.~C. Ellis, S.~Kuhlmann, K.~Kuehn, \emph{et~al.}, \enquote{Photonic ring resonator filters for astronomical oh suppression,} {\protect\JournalTitle{Opt. Express}} \textbf{25}, 15868--15889 (2017).

\bibitem{lindley2014demonstration}
E.~Lindley, S.-S. Min, S.~Leon-Saval, \emph{et~al.}, \enquote{Demonstration of uniform multicore fiber bragg gratings,} {\protect\JournalTitle{Optics express}} \textbf{22}, 31575--31581 (2014).

\bibitem{yu2024complete}
Q.~Yu, G.~Edvell, L.~Luo, and S.~G. Leon-Saval, \enquote{A complete design procedure of an aperiodic multichannel fbg for astronomical oh suppression,} {\protect\JournalTitle{Journal of Lightwave Technology}} \textbf{42}, 371--380 (2024).

\bibitem{ellis2015possibility}
S.~Ellis and J.~Bland-Hawthorn, \enquote{Possibility of observable signatures of leptonium from astrophysical sources,} {\protect\JournalTitle{Physical Review D}} \textbf{91}, 123004 (2015).

\bibitem{ellis2018astrophysical}
S.~C. Ellis and J.~Bland-Hawthorn, \enquote{Astrophysical signatures of leptonium,} {\protect\JournalTitle{The European Physical Journal D}} \textbf{72}, 1--9 (2018).

\bibitem{robertson2021seeking}
G.~Robertson, S.~Ellis, Q.~Yu, \emph{et~al.}, \enquote{Seeking celestial positronium with an oh-suppressed diffraction-limited spectrograph,} {\protect\JournalTitle{Applied Optics}} \textbf{60}, D122--D128 (2021).

\bibitem{yu2023aperiodic}
Q.~Yu, L.~Luo, G.~Edvell, \emph{et~al.}, \enquote{Aperiodic multi-notch fbg filters for astronomical positronium detection,} in \emph{AOPC 2023: Novel Technologies and Instruments for Astronomical Imaging and Spectroscopy,}  vol. 12965 (SPIE, 2023), pp. 92--100.

\bibitem{yu2024inverse}
Q.~Yu, B.~R. Norris, G.~Edvell, \emph{et~al.}, \enquote{Inverse design and optimization of an aperiodic multi-notch fiber bragg grating using neural networks,} {\protect\JournalTitle{Applied Optics}} \textbf{63}, D50--D58 (2024).

\bibitem{edvell2014optical}
G.~L. Edvell, \enquote{Optical structure writing system,}  (2014). US Patent 8,693,826.

\bibitem{Bryant2020}
J.~J. {Bryant}, J.~{Bland-Hawthorn}, J.~{Lawrence}, \emph{et~al.}, \enquote{{Hector: a new multi-object integral field spectrograph instrument for the Anglo-Australian Telescope},} in \emph{Ground-based and Airborne Instrumentation for Astronomy VIII,}  vol. 11447 of \emph{Society of Photo-Optical Instrumentation Engineers (SPIE) Conference Series} C.~J. {Evans}, J.~J. {Bryant}, and K.~{Motohara}, eds. (2020), p. 1144715.

\bibitem{DESI2016}
{DESI Collaboration}, A.~{Aghamousa}, J.~{Aguilar}, \emph{et~al.}, \enquote{{The DESI Experiment Part II: Instrument Design},} {\protect\JournalTitle{arXiv e-prints}} arXiv:1611.00037 (2016).

\bibitem{Leijtens2006}
X.~J. Leijtens, B.~Kuhlow, and M.~K. Smit, \enquote{Arrayed waveguide gratings,} in \emph{Wavelength filters in fibre optics,}  (Springer, 2006), pp. 125--187.

\bibitem{Gatkine2017}
P.~{Gatkine}, S.~{Veilleux}, Y.~{Hu}, \emph{et~al.}, \enquote{{Arrayed waveguide grating spectrometers for astronomical applications: new results},} {\protect\JournalTitle{Optics Express}} \textbf{25}, 17918 (2017).

\bibitem{Stoll2021}
A.~Stoll, K.~Madhav, and M.~Roth, \enquote{Design, simulation and characterization of integrated photonic spectrographs for astronomy ii: low-aberration generation-ii awg devices with three stigmatic points,} {\protect\JournalTitle{Opt. Express}} \textbf{29}, 36226--36241 (2021).

\bibitem{Gatkine:21}
P.~Gatkine, N.~Jovanovic, C.~Hopgood, \emph{et~al.}, \enquote{Potential of commercial sin mpw platforms for developing mid/high-resolution integrated photonic spectrographs for astronomy,} {\protect\JournalTitle{Appl. Opt.}} \textbf{60}, D15--D32 (2021).

\bibitem{Cvetojevic2012}
N.~Cvetojevic, N.~Jovanovic, J.~Lawrence, \emph{et~al.}, \enquote{Developing arrayed waveguide grating spectrographs for multi-object astronomical spectroscopy,} {\protect\JournalTitle{Opt. Express}} \textbf{20}, 2062--2072 (2012).

\bibitem{Cvetojevic2009}
N.~Cvetojevic, J.~S. Lawrence, S.~C. Ellis, \emph{et~al.}, \enquote{Characterization and on-sky demonstration of an integrated photonic spectrograph for astronomy,} {\protect\JournalTitle{Opt. Express}} \textbf{17}, 18643--18650 (2009).

\bibitem{vanWijk2020}
A.~van Wijk, C.~R. Doerr, Z.~Ali, \emph{et~al.}, \enquote{Compact ultrabroad-bandwidth cascaded arrayed waveguide gratings,} {\protect\JournalTitle{Opt. Express}} \textbf{28}, 14618--14626 (2020).

\bibitem{Hu2020}
Y.-W. {Hu}, S.~{Xie}, J.~{Zhan}, \emph{et~al.}, \enquote{{Integrated Arbitrary Filter With Spiral Gratings: Design and Characterization},} {\protect\JournalTitle{Journal of Lightwave Technology}} \textbf{38}, 4454--4461 (2020).

\bibitem{Blind2017}
N.~Blind, E.~L. Coarer, P.~Kern, and S.~Gousset, \enquote{Spectrographs for astrophotonics,} {\protect\JournalTitle{Opt. Express}} \textbf{25}, 27341--27369 (2017).

\bibitem{Dai2009}
D.~Dai, \enquote{Highly sensitive digital optical sensor based on cascaded high-q ring-resonators,} {\protect\JournalTitle{Opt. Express}} \textbf{17}, 23817--23822 (2009).

\bibitem{LeCoarer2007}
E.~Le~Coarer, S.~Blaize, P.~Benech, \emph{et~al.}, \enquote{Wavelength-scale stationary-wave integrated fourier-transform spectrometry,} {\protect\JournalTitle{Nature Photonics}} \textbf{1}, 473--478 (2007).

\bibitem{Bland-Hawthorn2011b}
J.~{Bland-Hawthorn}, J.~{Bryant}, G.~{Robertson}, \emph{et~al.}, \enquote{{Hexabundles: imaging fiber arrays for low-light astronomical applications},} {\protect\JournalTitle{Optics Express}} \textbf{19}, 2649 (2011).

\bibitem{Lewis2002}
I.~J. {Lewis}, R.~D. {Cannon}, K.~{Taylor}, \emph{et~al.}, \enquote{{The Anglo-Australian Observatory 2dF facility},} {\protect\JournalTitle{\mnras}} \textbf{333}, 279--299 (2002).

\bibitem{Croom2012}
S.~M. Croom, J.~S. Lawrence, J.~Bland-Hawthorn, \emph{et~al.}, \enquote{{The Sydney-AAO Multi-object Integral field spectrograph},} {\protect\JournalTitle{Monthly Notices of the Royal Astronomical Society}} \textbf{421}, 872--893 (2012).

\bibitem{Bryant2015}
J.~J. {Bryant}, M.~S. {Owers}, A.~S.~G. {Robotham}, \emph{et~al.}, \enquote{{The SAMI Galaxy Survey: instrument specification and target selection},} {\protect\JournalTitle{\mnras}} \textbf{447}, 2857--2879 (2015).

\bibitem{Bundy2015}
K.~{Bundy}, M.~A. {Bershady}, D.~R. {Law}, \emph{et~al.}, \enquote{{Overview of the SDSS-IV MaNGA Survey: Mapping nearby Galaxies at Apache Point Observatory},} {\protect\JournalTitle{\apj}} \textbf{798}, 7 (2015).

\bibitem{Bryant2011}
J.~J. {Bryant}, J.~W. {O'Byrne}, J.~{Bland-Hawthorn}, and S.~G. {Leon-Saval}, \enquote{{Characterization of hexabundles: initial results},} {\protect\JournalTitle{\mnras}} \textbf{415}, 2173--2181 (2011).

\bibitem{Bryant2014}
J.~J. {Bryant}, J.~{Bland-Hawthorn}, L.~M.~R. {Fogarty}, \emph{et~al.}, \enquote{{Focal ratio degradation in lightly fused hexabundles},} {\protect\JournalTitle{\mnras}} \textbf{438}, 869--877 (2014).

\bibitem{Wang2023}
A.~H. {Wang}, R.~{Brown}, J.~J. {Bryant}, and S.~{Leon-Saval}, \enquote{{Focal ratio degradation in optical fibres for the Hector integral field units},} {\protect\JournalTitle{\mnras}} \textbf{522}, 4310--4322 (2023).

\end{thebibliography}

\end{document}